\documentstyle[psfig]{laa}

\newcommand{\be}{\begin{equation}}
\newcommand{\en}{\end{equation}}
\def\ltsima{$\; \buildrel < \over \sim \;$}
\def\lsim{\lower.5ex\hbox{\ltsima}}
\def\loe{\lower.5ex\hbox{\ltsima}}
\def\gtsima{$\; \buildrel > \over \sim \;$}
\def\gsim{\lower.5ex\hbox{\gtsima}}
\def\goe{\lower.5ex\hbox{\gtsima}}

\def\aa #1 #2 {A\&A #1 #2}
\def\aass #1 #2 {A\&AS #1 #2}
\def\araa #1 #2 {ARA\&A #1 #2}
\def\mon #1 #2 {MNRAS #1 #2}
\def\apj #1 #2 {ApJ #1 #2}
\def\apjss #1 #2 {ApJS #1 #2}
\def\apjl #1 #2 {ApJ #1 #2}
\def\astrj #1 #2 {AJ #1 #2}
\def\nat #1 #2 {Nat #1 #2}
\def\pasj #1 #2 {PASJ #1 #2}
\def\pasp #1 #2 {PASP #1 #2}
\def\msai #1 #2 {Mem. SAIt #1 #2}
\def\ass #1 #2 {A.Sp.Sc. #1 #2}
\def\sci #1 #2 {Science #1 #2}
\def\phrevl #1 #2 {Phys. Rev. Lett. #1 #2}
\newcommand{\si}{\left(\frac{\Sigma}{0.005}\right)}

\newcommand{\ergs}{\rm \ erg \; s^{-1}}

\newcommand{\pdot}{ $\dot{P}_{orb}$ \su}
\newcommand{\befl}{ \vspace*{-17pt} \begin{flushright}}
\newcommand{\enfl}{\end{flushright}}

\def\mdot {\dot M}
\def\pdot {\dot P}

\def\cms  {\rm \ cm \, s^{-1}}
\def\gs   {\rm \ g  \, s^{-1}}

\def\msole {~M_{\odot}}

\title{The Neutron Stars of Soft X--Ray Transients}

\author{
S. Campana\inst{1,2}
\and M. Colpi\inst{3}
\and S. Mereghetti\inst{4} 
\and L. Stella\inst{5,2} 
\and M. Tavani\inst{6,4,2}
}

\begin{document}

\institute{
{Osservatorio astronomico di Brera, Via E. Bianchi 46, 
I-23807 Merate (Lc), Italy}
\and
{Affiliated to I.C.R.A.}
\and
{Dipartimento di Fisica, Universit\`a degli Studi di Milano,
Via Celoria 16, I-20133 Milano, Italy}
\and
{Istituto di Fisica Cosmica ``G.P.S. Occhialini'' 
del C.N.R., Via Bassini 15, I-20133 Milano, Italy}
\and
{Osservatorio astronomico di Monteporzio Catone, Via dell'Osservatorio
2, I-00040 Monteporzio Catone (Roma), Italy}
\and
{Columbia Astrophysics Laboratory, Columbia University, New York, 
NY 10027, USA}
}

\date{Received 4 November 1977; Accepted 20 April 1998}

\maketitle
\label{sampout}

\begin{abstract} 
Soft X--ray Transients (SXRTs) have long been suspected to contain old,
weakly magnetic neutron stars that have been spun up by accretion torques.
After reviewing their observational properties, we analyse the different 
regimes that likely characterise the neutron stars in these systems
across the very large range of mass inflow rates, from the peak of the 
outbursts to the quiescent emission.
While it is clear that close to the outburst maxima accretion onto the 
neutron star surface takes place, as the mass inflow rate decreases,
accretion might stop at the magnetospheric boundary because of the 
centrifugal barrier provided by the neutron star. For low enough mass 
inflow rates (and sufficiently short rotation periods), the radio pulsar 
mechanism might turn on and sweep the inflowing matter away.
The origin of the quiescent emission, observed in a number of SXRTs 
at a level of $\sim 10^{32}-10^{33}\ergs$, plays a crucial role 
in constraining the neutron star magnetic field and spin period.
Accretion onto the neutron star surface is an unlikely mechanism
for the quiescent emission of SXRTs, as it requires very low magnetic fields
and/or long spin periods. Thermal radiation from a cooling neutron star 
surface in between the outbursts can be ruled out as the only cause 
of the quiescent emission.

We find that accretion onto the neutron star magnetosphere and shock 
emission powered by an enshrouded radio pulsar provide far more plausible 
models.
In the latter case the range of allowed neutron star spin periods 
and magnetic fields is consistent with the values recently inferred
from the properties of kHz quasi-periodic oscillation in low mass 
X--ray binaries. If quiescent SXRTs contain enshrouded radio pulsars, 
they provide a missing link between X--ray binaries and millisecond pulsars.

\end{abstract}

\section{Introduction}

Transient X--ray sources are quiescent and undetected for most of the
time and undergo sporadic outbursts, typically lasting for 10--100~d,
during which they emit an intense X--ray flux. 
They were initially classified on the basis of their spectral
hardness, owing to the lack of a clear understanding of their nature
(Cominsky et al. 1978). The subsequent discovery of a number of
phenomena observed also in different classes of X--ray binaries (e.g.
X--ray pulsations and bursts; see e.g. White 1989) showed that there is a close
relationship between transient and persistent accreting compact
sources. White, Kaluzienski \& Swank (1984) introduced a revised
spectral classification that further extends this analogy. The spectra 
of {\it hard} X--ray transients (HXRTs) are characterised by 
equivalent temperatures $\gsim 15$ keV. These sources often contain 
a young, pulsating, neutron star orbiting a Be star companion and 
are clearly associated to persistent X--ray pulsars in high mass 
binaries (Maraschi, Treves \& van den Heuvel 1976). 
The outburst of {\it soft} X--ray transients (SXRTs), characterised 
by equivalent temperatures $\lsim 15$ keV, are often
accompanied by a pronounced increase in the luminosity of their
(faint) optical counterparts and by the onset of type I burst
(thermonuclear flashes on the surface of a neutron star) 
activity. These properties clearly associate SXRTs with low mass
X--ray binaries (LMXRBs) containing an old neutron star. The {\it
ultrasoft} X--ray transients are also associated to LMXRBs, but in
this case the similarity with the ``high state" spectra of persistent
black hole candidates (BHCs), together with the absence of bursts and
pulsations, suggests that these systems likely harbor a stellar mass
black hole (White \& Marshall 1984). This class has been later
extended to include also {\it hard tail} transient sources, based on
the analogy with the spectral characteristic of Cyg~X-1 in its ``low
state". 
The prediction on the nature of the compact object in different
classes of LMXRB transients has been brilliantly confirmed by mass
measurements which established A~0620--00, GS~2023+338,
GS~1124--68 and GRO J1655--40 as firm BHCs (McClintock \& Remillard 1986; 
Casares, Charles \& Naylor 1992;  Orsoz et al. 1996) and Cen X-4 as a 
neutron star (Shahbaz, Naylor \& Charles 1993).

Transients systems are characterised by an X--ray luminosity that
varies over many decades (variations between $10^{33}$ and
$10^{38}\ergs$ are not uncommon). Therefore they allow to investigate
accretion onto collapsed stars over a much larger range of
luminosities, and therefore accretion rates, than persistent sources.
This is well illustrated e.g. by the case of the HXRT
EXO~2030+375, a 42~s pulsator, that led to a remarkable
progress in the understanding of the physics of accretion onto
magnetic neutron stars (Parmar et al. 1989;  Parmar, White \& Stella
1989; Angelini, Stella \& Parmar 1989).
 
SXRTs, while still poorly studied, provide a unique opportunity to
gain crucial insights in the neutron stars that are hosted in
non-pulsating LMXRBs. 
According to current evolutionary scenarios, the neutron stars in
transients as well as persistent non-pulsating LMXRBs are gradually
spun-up by accretion torques to limiting periods ranging from
milliseconds to tens of milliseconds, depending on the value of their
residual magnetic field (Alpar et al. 1982; Bhattacharya \& van den
Heuvel 1991; Phinney \& Kulkarni 1994). Once accretion from the
companion star stops, the neutron stars in these systems are expected
to shine as ``recycled" radio pulsars orbiting a low mass companion.
These LMXRBs therefore likely represent the progenitors of the weak
magnetic field ($10^8-10^9$~G) millisecond radio pulsars (MSPs) that
are discovered in increasing number in the Galaxy and globular
clusters (Backer et al. 1982; Manchester et al. 1991; Taylor,
Manchester \& Lyne 1993). 

Despite numerous searches, fast coherent X--ray pulsations
in the persistent emission of LMXRBs directly arising from the neutron star
rotation have proved elusive (see e.g. Vaughan et al. 1994 and
references therein). Only after the launch of the Rossi X--Ray Timing
Explorer (RossiXTE) kiloHertz (kHz) quasi-periodic oscillations (QPOs)
have been discovered in several objects (for a review see van der Klis 
1997, 1998). In the great majority of the brightest X--ray binaries 
two kHz QPOs (between 0.3 and 1.2~kHz) have been discovered.  
These sources are usually classified according to the track in the 
colour-colour diagram for different luminosities (e.g. Hasinger \& van 
der Klis 1989; van der Klis 1995).
Atoll sources (or suspected) usually maintain their frequency difference 
($\sim 300-500$ Hz) constant over large variations of QPOs centroid 
frequencies and the centroid frequency appears to be positively 
correlated with the X--ray luminosity (4U 1728--34 Strohmayer 
et al. 1996; 4U 0614+091 Ford et al. 1997; KS 1731--260 Wijnands \& van 
der Klis 1997; 4U 1636--53 Wijnands et al. 1997; 4U 1735--44 Wijnands et 
al. 1998a; 4U 1820--30 Smale, Zhang \& White 1997; 4U 1705--44 Ford, van der Klis 
\& Kaaret 1998). 
A different case is presented by 4U 1608--52 for which a varying 
peak separation has been observed (Mendez et al. 1998).

In the Z source Sco X-1 (van der Klis et al. 1997) the frequency separation 
between the kHz QPOs decreases with mass accretion rate, but in the 
other Z sources (GX 5--1 van der Klis et al. 1996; Cyg X-2 Wijnands 
et al. 1998b; GX 340+0 Jonker et al. 1998; GX 17+2 Wijnands et al. 
1998c) the separation remains approximately 
constant, although a similar decrease in peak separation as found 
in Sco X-1 can not be excluded. 
 
Nearly constant pulsations at this frequency difference
have also been revealed during X--ray bursts in 4U 1728--34 
(2.8 ms; Strohmayer et al. 1996) and during a 
persistent emission interval of 4U 0614+09 (3.1 ms; Ford et al. 1997).
In the case of KS 1731--260 and 4U 1636--53 the frequency difference 
between the two QPO peaks is consistent with half the frequency of
the nearly periodic signals at $\sim 524$ and 581 Hz, respectively, 
that have been detected during type I bursts from these sources
(KS 1731--260: Smith, Morgan \& Bradt 1997; Wijnands \& van der Klis 1997; 
4U 1636--53: Zhang et al. 1996a; Wijnands et al. 1997).
In Aql X-1 a similar coherent modulation during an X--ray burst 
has been observed at $\sim 550$ Hz, whereas only one kHz QPO has been 
revealed at 750--830 Hz (Zhang et al. 1998).

The most straightforward interpretation of these findings is based on 
magnetospheric beat-frequency models (Alpar \& Shaham 1985; Lamb et al. 1985;
Miller, Lamb \& Psaltis 1998): the nearly coherent signal during type I bursts 
corresponds to the spin frequency of the neutron star, whereas the 
higher frequency kHz QPO arise from the Keplerian motion of matter in 
the innermost accretion disk region, close to the magnetospheric boundary 
or at the sonic radius. 
The lower frequency kHz QPO originates instead from modulated accretion 
at the beat frequency between the neutron star spin frequency and 
the Keplerian frequency at the magnetospheric boundary.

These kHz QPOs provide the first evidence that the neutron
stars in LMXRBs are spinning at periods of the order of milliseconds
and are the likely progenitors of MSPs. 

The large accretion rate variations that are characteristic of SXRTs
should allow the exploration of a variety of different regimes for 
the neutron stars in these systems which are unaccessible to persistent 
LMXRBs. 
While it is clear that, when in outbursts, SXRTs are powered by accretion, 
the origin of the low luminosity X--ray emission that has been detected
in the quiescent state of several SXRTs is still unclear.
An interesting possibility is that a MSP be visible in the quiescent 
state of SXRTs (Stella et al. 1994;  hereafter Paper I). This would provide a
``missing link" between persistent LMXRBs and recycled MSPs. 

This paper concentrates on various aspects of the physics of the
neutron stars in SXRTs. A short account of our main original results
has been given in Paper I. After a
review of the observations of SXRTs\footnote{Readers who are
familiar with the subject may refer to Table I only.} (Section 2, see also 
Tanaka \& Shibazaki 1996), we give a brief description of the models for 
the outburst mechanism (Section 3). 
In Section 4 we explore the different regimes that are expected for the 
neutron stars of SXRTs in the decay phase of their outbursts.
In Section 5 we expand on the different emission mechanisms which might be
responsible for the quiescent luminosity.  The conditions under which
the neutron stars of SXRTs evolve towards and remain within 
the radio pulsar region of the magnetic field -- spin period ($B-P$) diagram
are discussed in Section 6. The main conclusions of the 
paper are presented in Section 7.

\section {The properties of Soft X--ray Transients}
 
SXRTs are a fairly inhomogeneous class\footnote{Some sources
(especially in the vicinity of the Galactic Center) have been
classified as transients even if their peak luminosity was only a few
times higher than the instrumental detection limit. In the absence of
additional evidence for their transient behaviour, these sources
should be considered only as variable X--ray sources.}. 
Often their outbursts consist of a flux increase lasting a few days
that reaches X--ray luminosities of $L_X \sim 10^{37}-10^{38}\ergs$,
followed by a slower, nearly exponential decay with a timescale of
weeks to months. Outbursts of this kind have been observed in Cen
X-4, Aql X-1, 4U 1730--22 and A 1742--289 (White, Kaluzienski \& Swank 1984 
and references therein) and perhaps in a few other cases (see Table I).
Other sources (like EXO~0748--676, 4U~2129+47 and 4U~1608--52),
alternate long periods of relatively high (and often variable) X--ray
flux with others in which they are detected at a much lower level, if
at all. Unfortunately, the transitions between these intensity states
have been poorly studied so far.  With the possible exception of Aql X-1
(see below), the intervals between outbursts are irregular, often in
the 1--10 yr range, but for many sources only one outburst has
been observed so far. 

Available X--ray data on SXRTs are still sparse: in most cases
the outburst monitoring has been carried out with wide field
instruments, characterised by limited effective area, energy range
and, especially, sensitivity.
A relatively small number of pointed observations close to the outburst 
maxima have been obtained mainly with large area collimated detectors for
a few systems, whereas the quiescent emission, months to years away from 
the outbursts, has been investigated for $\sim 10$ SXRTs,
mainly with low energy X--ray telescopes. 
At least two SXRTs, Aql~X-1 and Cen~X-4, have been determined to emit
very different spectra in different states. While close to the outburst peak
the spectra are relatively soft (equivalent thermal
bremsstrahlung temperatures of $k\,T_{\rm br} \sim 5$~keV), at
intermediate luminosities ($\sim 10^{35}-10^{37}\ergs$) during the
rise and decay of the outburst, a high energy tail extending to at
least $\sim 100$~keV is detected (a similar tail is also seen in a
few persistent burst sources; e.g. Barret \& Vedrenne 1994). 
The evolution at the end of the outburst from luminosities of $\sim
10^{34}-10^{36}\ergs$ to quiescence is basically unknown. Only 
recently, BeppoSAX observations allowed to explore this luminosity 
range in the case of Aql X-1 (Campana et al. 1998; see Section 2.2.1).
The X--ray spectrum of Aql X-1 consists again of a soft component plus a
hard energy tail, similar to the one observed at higher luminosities.
The spectra in the quiescent state ($L_X \sim 10^{32}-10^{33}\ergs$)
are characterised by a soft component (equivalent black body temperatures of
$k\,T_{\rm bb}\sim 0.1-0.3$~keV). For those SXRTs observed above 
a few keV, a power-law like high energy tail has been revealed which, 
in the case of Aql X-1, hardens as the luminosity decreases below
$\sim 10^{33}\ergs$ (Campana et al. 1998).

Type I X--ray bursts have been observed in the active phase of thirteen
SXRTs. These burst of X--ray radiation are most likely due to thermonuclear 
flashes at the surface of accreting neutron stars (Maraschi \& Cavaliere 1977).
There is no evidence that these bursts differ in any property
from those of persistent LMXRBs (e.g. Lewin, van Paradijs \& Taam
1993, 1995), thus supporting the idea that SXRTs contain weakly magnetic
neutron stars accreting from a low mass companion. 

Up to now three SXRTs (4U 1608--52 Berger et al. 1996; KS 1731--260
Wijnands \& van der Klis 1997; Aql X-1 Zhang et al. 1998) have displayed 
kHz QPOs, when their X--ray luminosity was at a level of 
$10^{36}-10^{37}\ergs$.
In the case of 4U 1608--52 the centroid frequency of these QPOs 
(between 600 and 1100 Hz) does not correlate with the observed 
X--ray luminosity. 
For Aql X-1 a single kHz QPO in the same frequency range (750--830 Hz) has been 
observed at two different luminosities ($1.2-1.7\times 10^{36}\ergs$).
The persistent emission of KS 1731--260 shows twin kHz QPO around
900 and 1160 Hz, respectively. The frequency difference ($\sim 260$ Hz)
between these QPOs is consistent with half the frequency of the nearly 
periodic 524 Hz signal observed in a Type I burst from this source
(Smith, Morgan \& Bradt 1997; Wijnands \& van der Klis 1997).
When interpreted in terms of beat-frequency models (Alpar \& Shaham 1985; 
Lamb et al. 1985; Miller, Lamb \& Psaltis 1998) these results imply 
a neutron star spin period of $\sim 3.8$ ms.

The X--ray outbursts of SXRTs are often accompanied by a considerable
enhancement of their optical luminosity (optical novae). Increases up
to $\sim 6$ magnitudes with respect to the quiescent state have been
measured, which have greatly helped in the identification of the
optical counterparts of seven SXRTs. The optical spectra during outbursts
are usually characterised by a rather flat continuum with emission
lines (Balmer, He II, N III; van Paradijs \& McClintock 1995),
similar to those of bright persistent LMXRBs. These spectra result
mainly from reprocessing of the high energy photons at the accretion disk
and companion star. In some cases, the intrinsic spectrum of
the companion star becomes detectable in the quiescent state,
therefore making detailed photometric and spectroscopic
measurements possible. These studies have shown that SXRTs contain late type
stars (G or K), and in some cases allowed a determination of the
orbital periods. These are known for seven SXRTs and are in the range
from 4 to 19 hr, similar to those of LMXRBs (van Paradijs \&
McClintock 1994). Only for one SXRT (Cen X-4) the mass function has
been measured (Shahbaz, Naylor \& Charles 1993).

Radio emission has been observed during the outbursts of A1742--289
(Davis et al. 1976), Aql X-1 (Hjellming, Han \& Roussel-Dupr\'e 1990)
and Cen X-4 (Hjellming et al. 1988). 

SXRTs are located in the galactic plane with a distribution similar
to that of LMXRBs (van Paradijs \& White 1995). The presence of SXRTs
in globular clusters is of particular interest, due to their possible
evolutionary link with recycled MSPs. It is possible that some of the
dim ($L_X\lsim 10^{34}\ergs$) X--ray sources in globular clusters are
SXRTs in quiescence (e.g. Verbunt et al. 1994a; for an
alternative explanation see e.g. Grindlay 1994).

In Table I we summarise the main properties of the presently known or
suspected SXRTs. In the following we give a brief outline of the best studied
objects. In the compilation of Table I we have included all the transients for
which there are indications that they consist of a neutron star with
a low mass companion, excluding ultrasoft X--ray transients which
likely harbor a black hole.
Sources which are sometimes defined as transients in the literature,
but for which there is no clear evidence of flux variations greater
than a factor of a $\sim 100$, have not been included. Among these are:
KS 1732--273, EU 1737--132, EXS 1737.9--2952, GRS 1741.9--2853, GS 1826--24.

\setcounter{figure}{0}

\begin{figure*}[!p]
\psfig{figure=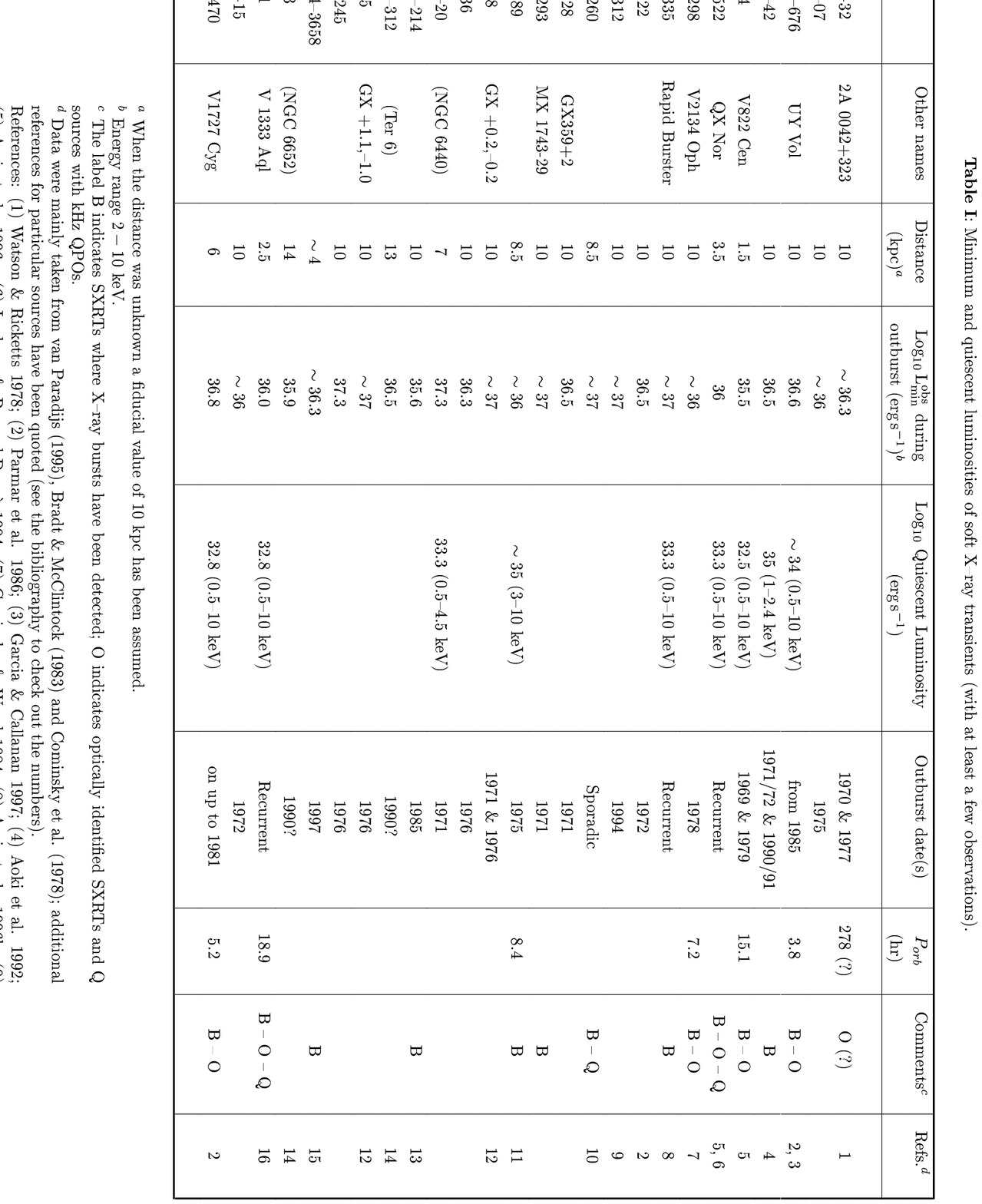}
\end{figure*}

\subsection{Soft X--ray Transients with fast rise and exponential decay
outbursts}
 
\subsubsection {Aql X-1}
 
Aql X-1 (4U~1908+005) is the most active SXRT known: more than 30
X--ray and/or optical outbursts have been detected. This 
led to several attempts to correlate the properties of different outburst
and to look for possible (quasi-)periodicities in the
recurrence times. There is evidence that the peak intensity of an
outburst correlates with the elapsed time from the previous one
(White, Kaluzienski \& Swank 1984; Kitamoto et al. 1993). A recurrence time of
$\sim 125$ d was quite evident in the 1969--1979 observations from
the Ariel V and Vela 5B satellites (Priedhorsky \& Terrell 1984).
However, this periodicity did not extend to the time of more recent
Ginga and optical observations (1987--1992), which on the contrary
suggest of a $\sim 310$ d periodicity (Kitamoto et al. 1993).
The outbursts of Aql X-1 are generally characterised by a fast rise
(5--10 d) followed by a slow exponential decay, with an $e-$folding
time of 30--70 d. Type I X--ray bursts were first discovered by
Koyama et al. (1981) during the declining phase of an outburst. A
periodicity of 132 ms, which persisted for only $\sim 1$ min, was
detected during the peak of a type I burst observed with the
Einstein SSS (Schoelkopf \& Kelley 1991). 

For a distance of $d\sim 2.5$~kpc, the corresponding 1--10~keV
luminosity is $L_X \sim (0.9-4)\times 10^{37}\ergs$. Close to the
outburst maxima the X--ray spectrum is soft with $k\,T_{\rm br} \sim
4-5$~keV. 
X--ray observations of Aql X-1 during the decay of an outburst have been 
collected in the $L_X\sim 10^{34}-10^{36}\ergs$ luminosity range. 
During the 1979 outburst Einstein MPC observed several times Aql X-1
(Czerny, Czerny \& Grindlay 1987).
The 1.2--10 keV spectrum when the source was at a level of a few
$10^{36}\ergs$ is well fit by a thermal bremsstrahlung model 
with the same temperature as in outburst. At a level of $\sim 
10^{35}\ergs$ the spectrum cannot be fit with the same model, but 
is instead consistent with a power-law model with photon index 
$\Gamma\sim 2.3$.
The same power-law spectrum was recovered about half a year later 
when the source was at a level $\sim 2\times10^{34}\ergs$ 
(Czerny, Czerny \& Grindlay 1987).
The ROSAT PSPC spectra during the 1990 and 1992 outbursts when 
$L_X\sim 10^{35}-10^{36}\ergs$ (0.1--2.4
keV) could not be fit satisfactorily by single component models
(Verbunt et al. 1994b). Finally the ASCA spectrum (0.5--10 keV) when the 
Aql X-1 luminosity was $\sim 2\times 10^{35}\ergs$ was well fit by 
a single power-law with photon index $\Gamma\sim 2$ (Tanaka \& Shibazaki 1996; 
Tanaka 1994).
Moreover, several episodes of hard X--ray emission have been
discovered by BATSE during 1991--1994 (Harmon et al. 1996), when the
X--ray luminosity was about $\sim 4\times 10^{36}\ergs$. The X--ray
spectra were characterised by a power-law of $\Gamma\sim 2-3$ and
extending up to 100 keV.

Probably due to its closeness, Aql X-1 is one of a few 
SXRTs detected in quiescence. The ROSAT HRI and PSPC revealed Aql X-1 on
three occasions at a level of $\sim 10^{33}\ergs$ in the
0.4--2.4 keV range. During these observations the spectrum was very 
soft and could be well fit either with a black body model with a
temperature of $k\,T_{\rm bb}\sim 0.3$ keV, a thermal bremsstrahlung
with a temperature of $k\,T_{\rm br}\sim 0.8$ keV or a power-law with
$\Gamma \sim 3$. The derived black body temperature implies
an emitting radius of $\sim 10^{5}$ cm (Verbunt et al. 1994b).

An outburst from Aql X-1 reaching a peak luminosity of $\sim 10^{37}\ergs$ 
(2--10 keV) was discovered (Levine et al. 1997) and monitored starting from 
mid-February, 1997 with the RossiXTE All Sky Monitor (see Fig. 1).
Several pointed observations were successively carried out leading to the
discovery of a nearly coherent modulation at 
$\sim 550$~Hz during a type I X--ray burst and a single QPO peak, with 
a frequency ranging from $\nu_{QPO}\sim 750$ to 830 Hz, at two different 
luminosities of $1.2-1.7\times10^{36}\ergs$ (Zhang et al. 1998).
 
Observations carried out with the BeppoSAX Narrow Field Instruments (NFIs) 
starting from March 8$^{th}$, 1997, allowed to study the final stages 
of the outburst decay  (see Fig. 1; Campana et al. 1998). 
At the time of the first BeppoSAX observation (which started on 
March 8$^{th}$, 1997) the source luminosity was decreasing very rapidly, 
fading by about 30\% in 11 hr, from a maximum level of $\sim 
10^{35}\ergs$. 
The second observation took place on March 12$^{th}$, 1997 when 
the source, a factor of $\sim 50$ fainter on average, reduced its flux 
by about 25\% in 12 hr. 
In the subsequent four observations the source luminosity attained its 
constant value of $\sim 6\times10^{32}\ergs$ (0.5--10 keV).
The sharp decrease after March 5$^{th}$ 1997 is well described by an 
exponential decay with an $e-$folding time $\sim 1.2$ d (see Fig. 1). The 
quiescent luminosity is consistent with the value previously measured  
with other satellites (e.g. Verbunt et al. 1994b).
 
The X--ray spectra during the fast decay phase, as well as that obtained 
by summing up all the observations pertaining to quiescence, could be fit 
with a model consisting of a black body plus a power-law.
The soft black body component remained nearly constant in temperature 
($kT_{\rm bb} \sim 0.3-0.4$ keV), but its radius decreased by a factor of 
$\sim 3$ from the decay phase to quiescence. The equivalent radius 
in quiescence ($R_{\rm bb}\sim 10^5$~cm) was consistent with the ROSAT 
results.
The power-law component changed substantially from 
the decay phase to quiescence: during the decay the photon index was 
$\Gamma \sim 2$, while in quiescence it hardened to $\Gamma\sim 1$.

The optical counterpart of Aql X-1 was identified in 1978 
with the variable K1IV star V1333 Aql (Thorstensen, Charles \& Bowyer 1978;
Shahbaz et al. 1996, 1997). Its quiescence magnitude is V=19.2
mag. Brightenings of up to $\sim 5$ mag have been observed during the 
X--ray outbursts. Chevalier and Ilovaisky (1991, 1997) monitored different 
optical outbursts of the source and determined a photometric orbital 
period of 18.9 hr.

Aql X-1 was observed at radio wavelengths with the VLA at 0.4 mJy (at 8.4 GHz)
during an outburst (Hjellming et al. 1990).  
This source was also searched during the quiescent phase
for pulsed radio emission at 400 MHz with the Jodrell Bank telescope.
This provided an upper limit of $\sim 10$ mJy during June 1989 (Biggs, Lyne \& 
Johnston 1989) and $\sim 3$ mJy during July--August 1989 (Biggs \& Lyne
1996). 

\begin{figure*}[!t]
\psfig{figure=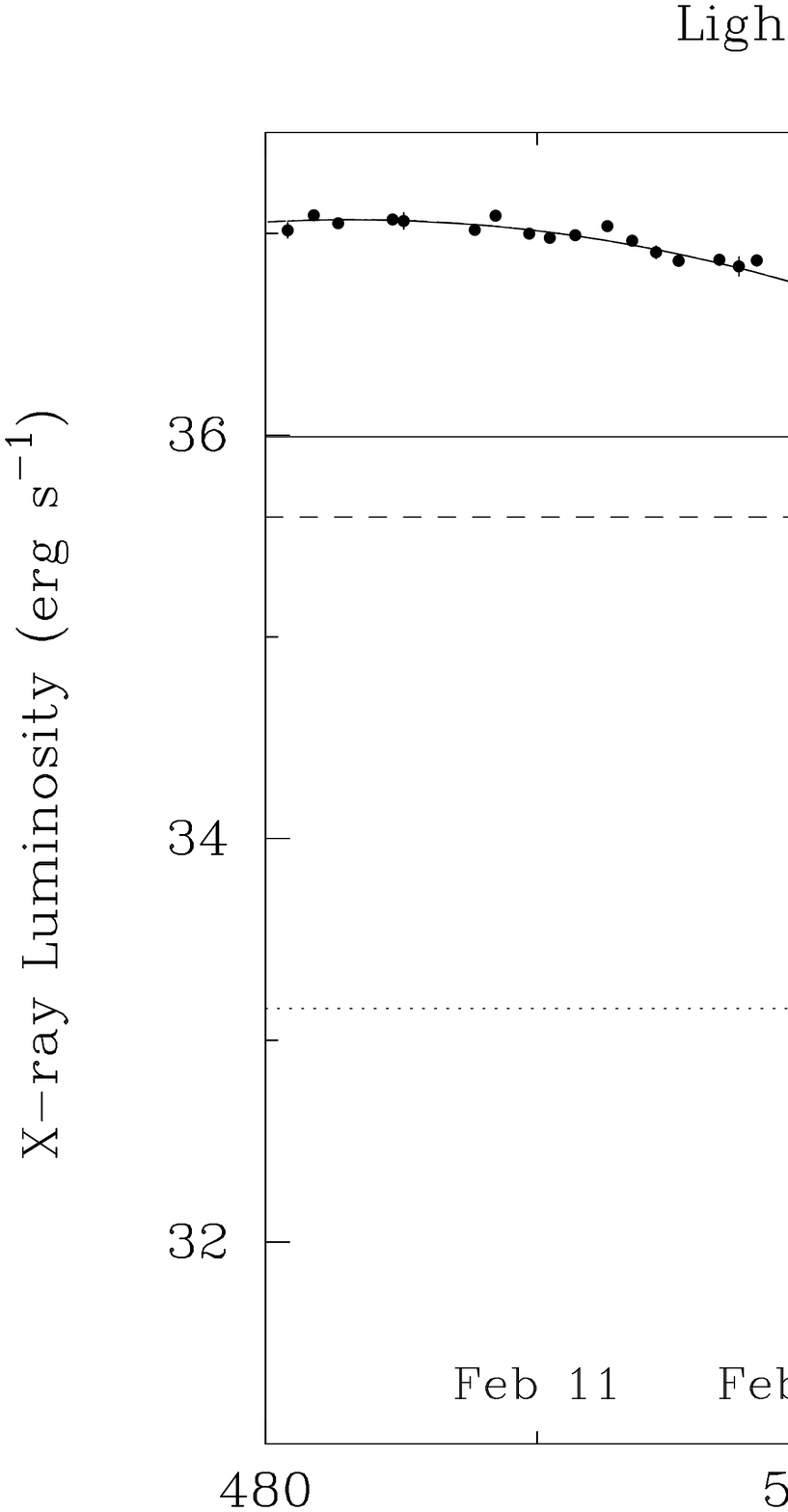,width=7cm}
\caption{
Light curve of the Feb.-Mar. 1997 outburst of Aql X-1.
Data before and after MJD 50514 were collected with the RossiXTE ASM
(2--10 keV) and the BeppoSAX MECS (1.5--10 keV), respectively.
RossiXTE ASM count rates are converted to (unabsorbed) luminosities using
a conversion factor of $4\times 10^{35}\ergs$ (before MJD 50512) and
$2\times10^{35}\ergs$ (after MJD 50512) as derived from RossiXTE spectral fits
(Zhang, Yu \& Zhang 1998).
BeppoSAX luminosities are derived directly from the spectral data (Campana
et al. 1998).
The evolution of the flux from MJD 50480 to MJD 50512 is well fit
by a Gaussian centered on MJD 50483.2. This fit however does not provide
an acceptable description for later times (see the dot-dashed line), not even
if the accretion luminosity is calculated in the propeller regime
(dashed line).
The straight solid line represents the X--ray luminosity corresponding to
the closure of the centrifugal barrier $L_{\rm min}$
(for a magnetic field of $10^8$ G and a spin period of 1.8 ms)
and the straight dashed line the luminosity gap due to the action of the
centrifugal barrier, $L_{\rm cor}$. The dotted line marks the minimum 
luminosity in the propeller regime ($L_{\rm lc}$).
} 
\label{fig_aql}
\end{figure*}

\subsubsection {Cen X-4}
 
Only two X--ray outbursts have been detected from Cen X-4. The second
outburst was observed in 1979, ten years after the first one (Conner,
Evans \& Belian 1969). Their rise times were $\sim 5-7$~d and
$e-$folding decay times of $\sim 30$~d. 
During the 1979 outburst Cen X-4 reached a peak flux of $\sim
5$~Crab, corresponding to $L_X\sim 4\times 10^{37}\ergs$ for $d \sim
1.2$~kpc (Kaluzienski, Holt \& Swank 1980). Type I bursts were
observed at intermediate luminosity levels (Matsuoka et al. 1980). A
variable, hard spectral component extending up to $\sim 100$ keV and
with an equivalent bremsstrahlung temperature $k\,T_{\rm br}$ of
30--70~keV was also revealed.  This component appeared slightly
before the outburst maximum, with a flux similar to that in the 3--6
keV range, and during the decay phase, with a factor of $\sim 5$
lower flux than the 3--6 keV flux (around $L_X\sim 2\times
10^{37}\ergs$; Bouchacourt et al. 1984).  The 1979 outburst was
particularly well studied and led to the identification of the
optical counterpart which had brightened by $\gsim 6$ magnitudes
(Canizares, McClintock \& Grindlay 1979).

ASCA detected Cen X-4 during quiescence at a level of $2.4\times
10^{32}\ergs$ (0.5--10 keV; Asai et al. 1996a).  The X--ray spectrum
was well fit by a black body component ($k\,T_{\rm bb}=0.16$ keV) plus an
additional power-law component with $\Gamma \sim 2-3$. The flux
from the two spectral components is comparable. The equivalent radius
of the black body emission is $\sim 1.8\times10^5$ cm, substantially 
smaller than the radius of a neutron star. A search for
X--ray pulsations gave negative results, providing an upper limit to
the pulsed fraction of $\sim 50\%$ between 8 ms and 8200 s (Asai et
al. 1996a).  During quiescence Cen X-4 was also observed with the
Einstein IPC (in 1980, $\sim 440$ d after the 1979 outburst; Petro et
al. 1981) and EXOSAT CMA (in 1986, van Paradijs et al. 1987).
Assuming the ASCA spectrum, Campana et al. (1997) found that both
measurements are consistent with the same value of the X--ray
luminosity derived with ASCA. A Ginga observation in 1991 provided an
upper limit of $\sim 5\times 10^{32} \ergs$ in the range 2--7 keV
(for a thermal brems\-strahlung spectrum with $T=5$ keV; Kulkarni et
al. 1992). A ROSAT HRI observation in 1995 revealed Cen X-4 at a
level comparable to that measured by ASCA, but showed a factor of 3 flux
variability in a few days (Campana et al. 1997).

Cen X-4 is one of the best studied SXRTs at optical wavelengths.
Extensive spectroscopic and photometric measurements of the optical
counterpart in quiescence (V=18.7 mag) led to the determination of
the orbital period (15.1 hr; Chevalier et al. 1989) and the mass
function ($\sim 0.2 \msole$, converting to a neutron star mass between
$0.5-2.1 \msole$;  Shahbaz, Naylor \& Charles 1993). The optical spectrum shows
the characteristics of a late K main sequence star, contaminated by
lines and continuum emission probably resulting from an accretion
disk (see, e.g., Cowley et al. 1988; Chevalier et al. 1989). The
exact nature and evolutionary state of the companion star, together
with the mechanism responsible for the mass transfer are the subject
of a long debate. The most likely scenarios involve either a peculiar
subgiant or a stripped giant, filling its Roche lobe, and viewed at
low inclination (McClintock \& Remillard 1990; Shahbaz, Naylor \& Charles 1993).
 
Radio emission had been revealed a few days after the 1979 outburst
at a level of $\sim 10$ mJy at both 1.5 and 4.8 GHz (Hjellming et al.
1988).  The quiescent phase of Cen X-4 was observed in the radio band
with the VLA, searching for pulsations and/or continuum emission.
Radio emission was not detected with an upper limit of
0.4 mJy at 1.4 GHz (Kulkarni et al. 1992).
 
\subsubsection {4U 1730--22}
 
The only known outburst from this source was observed with Uhuru in
1972 (Forman et al. 1978). The decay of the X--ray flux was
characterised by an $e-$folding time of $\sim 30$ d with evidence of a
secondary maximum. The thermal bremsstrahlung-like spectrum, with a 
temperature of $\sim 4$ keV, clearly indicated that this source 
belongs to the SXRT class (Cominsky et al. 1978).
 
\subsubsection {A 1742--289}
 
This SXRT, located close to the Galactic Center direction,
underwent a strong outburst in 1975, characterised by a fast rise
time to a peak flux of $\sim 2$~Crab (corresponding to a luminosity
of $\sim 4\times10^{38}\ergs$ at 8.5 kpc) followed by an exponential
decay with an $e-$folding time of 12~d (Branduardi et al. 1976). Though
the source lies in a very crowded region, the concomitant radio
outburst yielded an accurate position (Davis et al. 1976). 
ASCA has recently detected A~1742--289 at a 3--10 keV luminosity
ranging between $10^{35}-10^{36}\ergs$ (Maeda et al. 1996).
The source exhibited X--ray bursts and eclipses, yielding an orbital
period of 8.4 hr. 
The X--ray spectrum is highly absorbed ($N_H\sim 10^{23}$ cm$^{-2}$)
and could be well fit by a power-law with index 2.4 or a thermal
bremsstrahlung model with $k\,T_{\rm br}=7.5$ keV.  However, a
reanalysis of the Ariel V data from the 1975 outburst provided no
evidence for eclipses suggesting that the source detected by ASCA is
perhaps a new one, rather than the quiescent counterpart of A~1742--289
(Kennea \& Skinner 1996).

\subsection {Soft X--ray Transients with extended on/off periods}
 
The sources of this sample are characterised by on and off states 
lasting several months to years. Three out of four systems show partial 
X--ray eclipses and/or dips, indicating that they are viewed from a high
inclination. In some instances there is strong evidence that
the central X--ray source is hidden by the accretion disk and the
observed X--rays are scattered along our line of sight by an extended
photo-ionised corona above the disk (i.e. they are accretion disk corona
sources; White \& Holt 1982). A different case is presented by 4U
1608--52 which shows intermediate properties between this class of
SXRTs and those displaying outbursts with fast rise and nearly
exponential decay.

\subsubsection{EXO 0748--676}
 
EXO 0748--676 was discovered during a slew with the EXOSAT
satellite in 1985. The intensity decayed in the first 2 months after
the discovery, as expected for classical transients. However, in
June--July 1985 and January 1986 it was again in a bright state with
$L_X\sim 10^{37}\ergs$ (Parmar et al. 1986). 
The source was still active when reobserved with Ginga in 1989
(Parmar et al. 1991) and more recently with ASCA, even if at a lower
level (Corbet et al. 1994; Thomas et al. 1997).
Type I bursts (with photospheric radius expansion), dips and partial
eclipses at the 3.8 hr orbital period were observed with EXOSAT.
Parmar et al. (1986) found a quiescent X--ray luminosity of $10^{34}\ergs$
(0.5--10 keV), while Garcia \& Callanan (1998) infer from the same
data a black body temperature of $k\,T_{\rm bb}\sim 0.2$ keV.
Based on ASCA data Thomas et al. (1997) find evidence of a previously
unreported soft excess.
Optical observations, which detected EXO 0748--676 in the range
V$\,\sim 17.5-16.8$ mag also testify that the source remained active
during the period 1985--1993. Optical variability as well as optical
bursts resulting from reprocessed X--ray bursts have also been
observed.

\subsubsection{4U 2129+47}
 
4U 2129+47 shows partial X--ray eclipses at the orbital period of 5.2
hours, as well as type I bursts. Before 1984 it was considered a
persistent LMXRB, since it had been detected and studied with all the
major X--ray satellites. Its relatively low luminosity and smooth
orbital modulation at 5.2 hr in the active state ($\sim 10^{35}-10^{36}\ergs$
for $d\sim 6$ kpc)
suggests that 4U 2129+47 is an accretion disk corona source (White \&
Holt 1982). During this active state the optical counterpart was identified
(Thorstensen et al. 1979).
EXOSAT CMA observations in September 1983 provided only
upper limits corresponding to a luminosity of $\lsim 10^{34}\ergs$
(Pietsch et al.  1986). Subsequent X--ray and optical observations
showed that the source entered a long period of quiescence (Garcia et
al. 1989; Molnar \& Neely 1992) and led to its inclusion in the SXRT
group. 
A ROSAT HRI observation detected this source at a level of $\sim
3\times 10^{33}\ergs$ (0.3--2.4 keV; Garcia 1994). Garcia \& Callanan 
(1998) derived a 
quiescent luminosity of $6\times 10^{32}\ergs$ (0.5--10 keV) for a black 
body spectrum with a temperature $k\,T_{\rm bb}=0.2$ keV.
The quiescent X--ray light curve, obtained with the ROSAT HRI, 
does not show strong evidence for 
the partial eclipses characteristic of the active state, indicating 
that either the vertical extent of the disk is drastically reduced or 
that the disk is not present during quiescence (Garcia 1994).
Note that the optical spectrum of 4U 2129+47 during quiescence does not
display any characteristic feature of an accretion disk (Garcia 1994).

The companion was classified as a F9 subgiant, but no evidence was found of 
the expected ellipsoidal and radial velocity variations at the orbital period 
(Garcia et al. 1989).

An upper limit of $\sim 8$ mJy at 610 MHz for pulsed radio emission
has been set for 4U 2129+47 during quiescence (Biggs, Lyne \& Johnston 1989) and
of 2 mJy at 400 MHz (Biggs \& Lyne 1996); at 1.4 GHz the $4\,\sigma$
upper limit is of 0.25 mJy (Kulkarni et al. 1992).
 
\subsubsection{4U 1658--298}

4U 1658--298, was discovered in 1976 when an isolated Type I burst
was detected (Lewin, Hoffman \& Doty 1976a). The corresponding SXRT
was discovered only two years later during a strong
outburst reaching a flux of $\sim 10^{-9}\ergs$
cm$^{-2}$ (Lewin et al. 1978; Share et al. 1978). During this
outburst, X--ray eclipses and dips were also detected (Cominsky,
Ossmann \& Lewin 1983). Hard X--rays up to $\sim 80$ keV from this source 
were revealed with the A4 experiment on board HEAO-1 (Levine et al. 1984).

\subsubsection{4U 1608--52}
 
The persistent source 4U 1608--52 detected by Uhuru and OSO-7 in
1971--1973, and the transient observed with Ariel V in November 1975
and with Ariel V, SAS-3 and HEAO 1 in July--September 1977 were
recognised to be the same object (Fabbiano et al. 1978 and references
therein). The type I bursts (with photospheric radius expansion
indicating a 3.6 kpc distance) observed from this region in the Norma
constellation were also attributed to this source, thus providing the
first case of transient-burster association. The bursts from 4U
1608--52 were later confirmed with HAKUCHO (Murakami et al. 1980) and
EXOSAT (Penninx et al.  1989). 
 
The Vela 5B data detected 8 outbursts during 1969--1979 as well as a
persistent emission at a level of $\sim 10^{36}\ergs$ (Lochner \&
Roussel-Dupr\`e 1994). The outbursts are characterised by either a
sharp rise and an exponential decay or a much more symmetric
evolution.  
Like other SXRTs, 4U 1608--52 exhibited a spectral transition from a
thermal bremsstrahlung to a power-law like spectrum when the
luminosity decreased below $\sim 10^{37}\ergs$ (Mitsuda et al.
1989). A hard power-law like spectrum has also been detected with
BATSE during the 1991 outburst with a power-law slope of $\Gamma\sim 1.8$ 
and a steepening above $\sim 65$ keV (Zhang et al. 1996b). 
The high energy spectrum could be equally well fit by either a 
Sunyaev-Titarchuk Comptonisation model or a broken power-law.

In 1993 ASCA revealed 4U 1608--52 at a much lower X--ray luminosity of
$\sim 2\times10^{33}\ergs$ in the 0.5--10 keV energy range (Asai et al.
1996a).  The spectrum could be well fit by a black body with $k\,T_{\rm
bb}=0.30$ keV or a bremsstrahlung with $k\,T_{\rm br}=0.32$ keV. The
black body emission radius is $\sim 1.5\times10^5$ cm, substantially 
smaller than the radius of a neutron star.  A
periodicity search based on the ASCA light curves provided an upper
limit of 50\% rms in the range 8 ms--8200 s, which is valid only if
the orbital period is longer than 2 d (Asai et al. 1996a).
 
This source was active again in 1996 and was observed by RossiXTE at a level
of $\sim 2\times10^{37}\ergs$ with a $k\,T_{\rm br}\sim 5$ keV 
bremsstrahlung spectrum
(Marshall \& Angelini 1996). RossiXTE observations revealed also a
variable QPO feature at 850--890 Hz, the frequency variations of
which did not correlate with intensity changes (Berger et al. 1996).
A second kHz QPO has been recently revealed at about 1100 Hz simultaneous
with the 600--900 Hz peak (Yu et al. 1997) previously known (Mendez et 
al. 1998). There is evidence that the frequency separation varied between 
230--290 Hz, perhaps providing the first example of a variable kHz peak 
separation in an atoll source.

The optical counterpart, identified
during the 1977 outburst, is a reddened faint star ($I\sim 18.2$ mag)
which becomes fainter in quiescence ($I>20$ mag, $B>22$ mag; Grindlay \&
Liller 1978). This was re-discovered during the 1996 outburst about 
150~d after the peak at a level of $R=20.2$ mag and $J=17.2$ mag, when the 
source was still active in the X--rays (Wachter 1997).
One year before the outburst the source was detected at $J=18.0$ mag 
and $R>22$ mag (Wachter 1997).

\subsubsection{4U 1730--335: the Rapid Burster}

The Rapid Burster alternates periods
of activity lasting several weeks, to period of quiescence, during which 
the X--ray luminosity decreases by more than three orders of magnitude.
During the active periods the Rapid Burster emits a variety of
combinations of type I and II bursts, making it unique among LMXRBs
(Lewin et al. 1976b).
Here we do not describe in detail the very complex phenomenology of
this source and we refer to the review by Lewin, van Paradijs \& Taam
(1995) and references therein.

In the state in which the Rapid Burster shows the closest resemblance
to other SXRTs, namely persistent X--ray emission and type I X--ray
bursts, the average luminosity is about $10^{37}\ergs$ (for the $\sim 10$ 
kpc distance of the globular cluster Liller 1) and the
spectrum can be described by a thermal bremsstrahlung with $k\,T_{\rm
br}\sim 10$ keV (Barr et al. 1987).  During quiescence an upper limit
of $\sim 10^{34}\ergs$ was obtained with Einstein (Grindlay 1981).
Recently, the Rapid Burster has been detected in quiescence by ASCA
at a level of $3\times 10^{33}\ergs$ (Asai et al. 1996b). However, the ASCA
point spread function ($\sim 3'$) is comparable to angular radius
of the Liller 1 cluster ($3.3'$) and it cannot be ruled out
that the measured X--ray flux is due to other sources within the
cluster.

RossiXTE observations of X--ray bursts from the Rapid Burster 
did not lead to the discovery of kHz QPOs (Guerriero, Lewin
\& Kommers 1997).
A radio transient with flux density correlated with the RossiXTE ASM X--ray 
flux of the Rapid Burster has been recently observed (Rutledge et al.
1998).

\subsection{Soft X--ray Transients with poorly sampled outburst light curves}
 
\subsubsection{MX 0836--42}
 
This source was discovered at the end of 1971 with OSO-7 and Uhuru
(Markert et al. 1977; Cominsky et al. 1978) at a level of $8\times
10^{-9}\ergs$ cm$^{-2}$. The spectrum was soft, but owing to poor
coverage the light curve of the outburst could not be determined
accurately. MX 0836--42 remained undetected until a new period of
activity occurred in 1990--1991. Ginga detected a variable X--ray
flux from this source from the end of November 1990 until February
1991.  ROSAT observed an active state at a level of $\sim
10^{-10}\ergs$ cm$^{-2}$ (1.0--2.4 keV) around the middle of November 
1990 during the all sky survey and re-observed it at a level
$\sim 15$ times lower in May 1991. These observations led to the
discovery of Type I bursts (Aoki et al. 1992) and a substantial
reduction of its error box (Belloni et al. 1993), confirming the LMXRB
nature of this source.
  
\subsubsection{MX 1746--20}
 
Little is known about this transient which has been unambiguously
observed only in January 1972. It was discovered with the OSO-7
satellite (Markert et al. 1975) and tentatively identified with the
globular cluster NGC 6440. This association was also supported by the
more precise source location obtained with the Uhuru data (Forman,
Jones \& Tananbaum 1976). The peak X--ray luminosity was about $10^{37}\ergs$
for the distance of this cluster ($d=7$ kpc).
Cominsky et al. (1978) classified MX 1746--20 as a soft transient, on
the basis of a spectral fit yielding a thermal bremsstrahlung
temperature of $\sim 4$ keV.
 
It is generally assumed that the dim source detected with the Einstein HRI 
at a level of $\sim10^{33}\ergs$ in the globular cluster NGC~6440 (Hertz \&
Grindlay 1983) is the quiescent counterpart of MX 1746--20. ROSAT HRI
observations confirmed this detection at a similar level (Johnston,
Verbunt \& Hasinger 1995).

\subsection{Galactic Center transients}
 
Over the last few years several X--ray sources in the Galactic Center
region have been discovered with Ginga, TTM, ART-P and SIGMA (e.g.
in't Zand et al. 1989; Sunyaev et al. 1991). Little is known about
these sources, owing the lack of optical identifications, poor
temporal coverage and spectral information.  The maximum X--ray
luminosity observed from KS 1730--312, 4U 1735--28, EU 1737--132, GS
1741.1--2859 are of the order of $L_X\sim 10^{38} \ergs$. On the other hand
the highest detected luminosities of KS 1732--273, EXS 1737.9--2952,
GRS 1741.9--2853 are considerably lower ($L_X\sim 10^{37} \ergs$).
Some other sources were only a factor of a few above the instruments'
detection thresholds: among these are KS 1632--477, KS 1724--356, 
KS 1731--260, KS 1739--304, KS 1741--293. It is therefore unclear whether 
these sources are to be considered transient sources and, in particular,
SXRTs.
 
\subsubsection{KS 1731--260}

The best studied of the latter sources, KS 1731--260, shows only
modest X--ray luminosity variations, casting doubts on its inclusion
in the SXRT class.  The source was discovered by TTM on board
MIR-KVANT (Sunyaev 1989) at a level of $\sim 10^{37}\ergs$ (2--27
keV) for a distance of 8.5 kpc.  The spectrum was well fit by a
bremsstrahlung spectrum with $k\,T_{\rm br} \sim 6$ keV; X--ray bursts
were also observed (Sunyaev et al. 1990).  KS 1731--260 was also
detected by ART-P, SIGMA (revealing a hard energy tail extending up to 
energies of $\sim 40$ keV; Barret et al. 1992) and during the ROSAT all-sky
survey (see e.g. Smith, Morgan \& Bradt 1997). 
RossiXTE revealed a fairly coherent signal during a $\sim 2$ s
interval starting at the end of the contraction phase of an X--ray
burst. These oscillations likely indicate the presence of neutron star spinning
at 1.9 ms (Smith, Morgan \& Bradt 1997). Moreover, two simultaneous QPO peaks
around 900 and 1160 Hz have also been discovered with RossiXTE (Wijnands \& 
van der Klis 1997), the difference frequency of which is consistent with
half the frequency observed by Smith, Morgan \& Bradt  (1997), suggesting that the 
neutron star spin period is instead $\sim 3.8$ ms.
 
\section{Outburst models}

Models of SXRT outbursts closely parallel models of dwarf nova
outbursts.  One group of models involves accretion disk instabilities
(Cannizzo 1993 and references therein; Huang \& Wheeler 1989;
Mineshige \& Wheeler 1989; Cannizzo, Chen \& Livio 1995) and the
other instabilities in the mass transfer from the companion star
(Osaki 1985; Hameury, King \& Lasota 1986). For a recent review 
see Tanaka \& Shibazaki (1996).

In both approaches, a compact star is continuously accreting from a
disk fed by a companion star filling or almost filling its Roche lobe
and transferring matter at relatively low rate in quiescence.  In
disk instability models, the accretion disk can become locally
thermally and viscously unstable due to strong opacity variations
caused by partial ionisation of hydrogen (Cannizzo 1993). 
The mechanism of mass transfer instability instead relies on the slow
expansion of the superadiabatic convective layers of the companion
star which is irradiated by hard ($E\gsim 10$ keV) X--rays, supposedly
produced by low-level accretion onto the neutron star surface
(Hameury, King \& Lasota 1986). After the onset of accretion at a high rate,
the high energy radiation is intercepted by the evolving accretion
disk and the companion contracts again within a few weeks, leaving a
disk fed at a low rate. 

In a critical review Lasota (1995) pointed out the major problems of
these two classes of models: disk instability models fail to
reproduce the observed recurrence times, which are usually too short; 
mass transfer instability models require a quiescent hard X--ray luminosity 
at a level of $10^{34}-10^{35}\ergs$ and a subgiant or
stripped giant companion in order to produce outburst recurrences in
the observed range. 

In the framework of disk instability model van Paradijs (1996) recently
singled out the importance of X--ray heating of the accretion disk. 
Including these effects, the modified disk instability criterion 
leads to recurrence times in agreement with those of
SXRT outbursts. This supports the idea that the disk instability models
apply.

Independent of the specific accretion instability responsible for the
transient behaviour (which of course might be different from those 
outlined above), we focus our attention on the properties of
SXRTs in outburst and during quiescence, to explore the different
accretion regimes and to constrain the physical parameters of the
neutron stars.

\section {Accretion and other regimes in Soft X--ray Transients}
 
If the neutron stars of SXRTs possess a magnetic field of at least
$10^8-10^9$~G, as suggested by the evolutionary connection with MSPs
and by the magnetospheric interpretation of the kHz QPOs observed in 
several LMXRBs, then regimes ranging from accretion onto the magnetic
polar caps to radio pulsar activity are potentially accessible.
Depending on the spin period and magnetic field, transitions from one
regime to another can in principle occur during the outburst cycle,
as a consequence of drastic changes in the mass inflow rate.

When SXRTs are in outburst, the occurrence of type I bursts and kHz QPOs
as well as the similarity of their properties (especially the X--ray 
spectrum) with those of persistent LMXRBs testify that accretion down 
to the neutron star surface takes place.  
Matter inflowing from the inner Lagrangian
point will give rise to an accretion disk in the Roche lobe of the
compact object, due to its high angular momentum. The presence of
a disk during the outburst phase is also inferred from the
enhancement of the optical luminosity, the non-thermal 
continuum spectra and the presence of weak emission lines, which
strengthen during the decline, especially Balmer lines (van Paradijs \& 
McClintock 1995). 

If the neutron star is magnetic,
the accretion disk cannot extend to radii smaller than the
magnetospheric radius, $r_{\rm m}$, where the pressure of the 
magnetic dipole field ($P_{\rm mag}\propto \mu^2\,r^{-6}$, with $\mu$ the
magnetic dipole moment\footnote{In the following we use the magnetic dipole 
moment $\mu=B\,R^3/2$, where $B$ is the neutron star magnetic field and $R$
the neutron star radius.} and $r$ the distance from the neutron star)
balances the pressure of the incoming matter.  We consider here the
simpler case of spherical accretion (see also Paper I) and refer to appendix 
A for a more detailed treatment based on the Ghosh \& Lamb magnetically
threaded-disk model (1979a; 1979b; 1992). 
The magnetospheric radius in the spherical free-fall approximation is
given by:
\begin{eqnarray}
r_{\rm m}^{sp} & = & \Bigl({{\mu^4}\over{2\,G\,M\,\mdot^2}}\Bigr)^{1/7} 
\nonumber\\ & = & 
2\times 10^7\,\mdot_{15}^{-2/7}\,B_9^{4/7}\,
M_{1.4}^{-1/7}\,R_6^{12/7} ~{\rm cm} \label{rmsp}
\end{eqnarray}
\noindent where $\mdot=10^{15}\,\mdot_{15}\gs$ is the accretion rate, 
$B=10^9\,B_9$ G, $M=1.4\,M_{1.4}\msole$ and $R=10^6\,R_6$ cm are the 
neutron star magnetic field, mass and radius, respectively.
The inflow of matter can proceed toward the neutron star only if several
conditions are met.
Within $r_{\rm m}$ matter is enforced to corotate with the neutron star 
magnetosphere. X--ray pulsations are likely to occur if a sizeable fraction of
the material gets attached to the magnetic field lines and accretion
takes place preferentially onto the magnetic polar caps. Accretion
onto the neutron star surface gives rise to a luminosity of 
\be
L(R)=G\,M\,\mdot/R \label{princ} 
\en

The X--ray luminosity decay from the outburst peak takes place 
as a consequence of the decreasing accretion rate. Accretion onto the
neutron star surface can continue as long as the centrifugal drag
exerted by the corotating magnetosphere on the accreting material (in
almost Keplerian rotation) is weaker than gravity (i.e. the ``centrifugal
barrier" is open, e.g. Illarionov \& Sunyaev 1975; Stella, White \& 
Rosner 1986). On the contrary, accretion onto the
neutron star surface is inhibited when the magnetospheric radius becomes 
larger than the corotation radius, $r_{\rm
cor}=\bigl( {{G\,M\,P^2}\over {4\,\pi^2}}\bigr)^{1/3}$ (where $P$ is
the neutron star spin period), since the drag exerted
by the neutron star magnetic field becomes super-Keplerian.
When this occurs the system enters the ``propeller" phase
(i.e. the centrifugal barrier closes). 
Due to the scaling of $r_{\rm m}$ with $\mdot$ (the magnetospheric
radius expands as the accretion rate decreases), there exists a minimum 
mass inflow rate, $\mdot_{\rm min}$, below which the centrifugal barrier 
cannot be penetrated; this corresponds to an accretion luminosity of 
\be
L_{\rm min}(R)=G\,M\,\mdot_{\rm min}/R \label{lminn}
\en 
In Fig. 2 we sketch the dependence of the accretion
luminosity as a function of the inflow rate.
In the case of spherical accretion this equation becomes:
\be
L_{\rm min}^{sp}(R)=2\times 10^{36}\,B_9^2\,M_{1.4}^{-2/3}
\,R_{6}^{5}\,P_{-2}^{-7/3} ~{\rm erg\,s^{-1}} \label{lminsp}
\en
\noindent where $P_{-2}$ the spin period in $10^{-2}$ s
(note that Eq. 1 in Paper I is incorrect for a factor of 2). 
In the case of the Ghosh \& Lamb
disk (see Eq. \ref{LXGL}) this limiting luminosity is about a factor
of $\sim 10$ lower than for spherical accretion for the same set
of parameters and for a critical fastness parameter $\omega_c=0.5$ 
(see Appendix A). 
This is due to the smaller neutron star magnetosphere in the Ghosh \& 
Lamb disk case (a factor of $\sim 3$, see Eq. \ref{compa}).

\begin{figure*}[htb]
\centerline{\psfig{figure=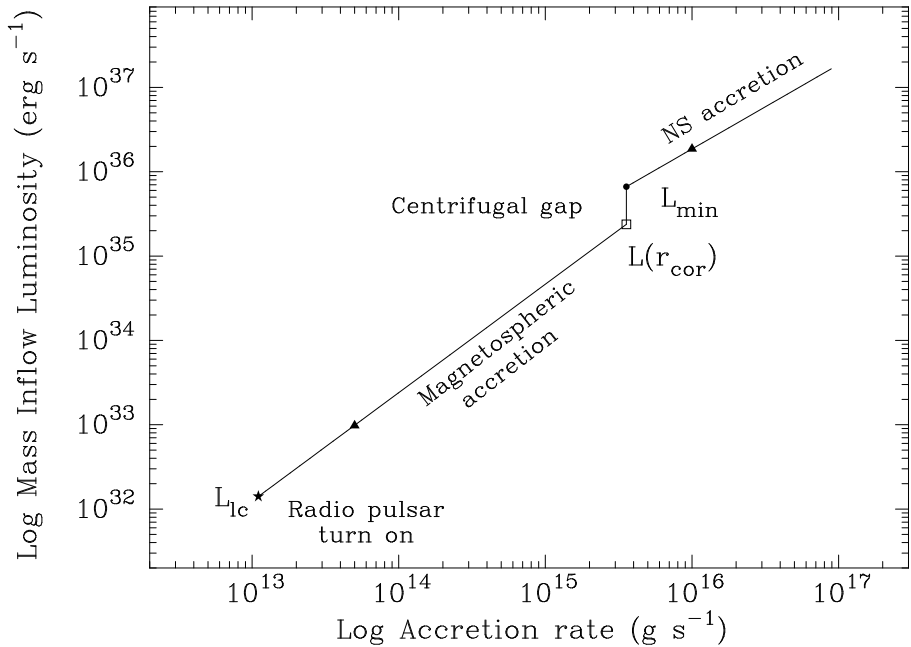,width=9cm}\hfill}
\caption{Different regimes of spherical accretion onto a 
rotating magnetic neutron star for
decreasing mass inflow rates (see arrows). A neutron star with spin 
period of 4 ms
and magnetic field of $10^8$ G has been considered. The upper line
describes accretion onto the neutron star surface ($L_X\propto\mdot$,
cf. Eq. \ref{princ}). The filled dot represents the onset of the
centrifugal barrier, corresponding to Eq. \ref{lminn}. The open
square marks the point where the centrifugal barrier is closed and 
a much smaller luminosity is released from accretion onto the 
magnetosphere (cf. Eq. \ref{L_min}).
In this regime the emitted luminosity is $L_X \propto
\mdot^{9/7}$ (cf. Eq. \ref{lrmsp}). For even lower mass inflow rates 
the radio pulsar mechanism can turn on (filled star, see text 
and Eq. \ref{L_lc}).} 
\label{propell}
\end{figure*}

Constraints on the neutron star spin period and magnetic field 
can be inferred for a SXRT by
comparing the minimum luminosity $L_{\rm min}(R)$ defined in Eq.
\ref{lminn} to the minimum observed X--ray luminosity $L_{\rm
min}^{obs}$ at the end of an outburst, as long as one can be confident
that this luminosity is still produced by accretion onto the neutron
star surface. For a given source, the minimum observed X--ray luminosity
produced by neutron star accretion defines a line in the usual magnetic
field -- spin period
($B-P$) diagram for radio pulsars: in order to accrete at the
observed rate, a neutron star must posses a spin period and magnetic
field such as to lie to the right of that line. These lines are
similar to the usual spin-up line which is drawn for $L=L_{\rm
Edd}=1.8\times 10^{38}\,M_{1.4}\ergs$. Lines of constant
$L^{sp}_{\rm min}$ are shown in Fig. 3 for different
luminosities.

Once the centrifugal barrier closes (i.e. when $r_{\rm m}\gsim r_{\rm
cor}$) the source enters the propeller regime. 
The accretion flow is
halted at the magnetospheric radius and an accretion luminosity of up to
\be
L(r_{\rm m})=G\,M\,\mdot_{\rm min}/r_{\rm m} \label{L_min}
\en
\noindent is released (see Fig. 2).
In the case of spherical accretion and just after the closure
of the centrifugal barrier (i.e. for $r_{\rm m}\sim r_{\rm
cor}$) this luminosity is
\be
L(r_{\rm cor})=2\times10^{35}\,B_9^2\,M_{1.4}^{-1}\,P_{-2}^{-3}
\,R_6^{6} {\rm \ erg\,s^{-1}} 
\label{lcor}
\en
In the case of disk accretion an extra factor of 1/2 is present on the 
right hand side of Eq. \ref{L_min}, due to the fact that, in general, 
the disk matter at $r_{\rm m}$ is expected to retain the kinetic
energy of its orbital motion (note that the magnetosphere rotates faster)
and therefore only 1/2 of the gravitational energy is released.
(For a Ghosh \& Lamb disk this luminosity is a factor 
of $\sim 25$ lower for the same set of parameters, see Appendix A).
This transition is expected to occur when the matter inflow rate
decreases below $\mdot_{\rm min}$ in the decay phase of an X--ray
transient outburst (Stella, White \& Rosner 1986); correspondingly a sharp
decrease of the X--ray luminosity should be observed.
Corbet (1996) has shown that the ratio of accretion luminosity
before and after the closure of the centrifugal barrier,
$\Delta$, does not depend on the neutron star magnetic field nor on
the accretion rate but only on its spin period (and radius):
\be 
\Delta={{L_{\rm min}(R)}\over{L(r_{\rm cor})}}=\Bigl({{G\,M\,P^2}
\over{4\,\pi^2}}\Bigr)^{1/3}\,{1 \over R}
\label{corbet}
\en
In the case of disk accretion an extra factor of 1/2 is present (see above).

Evidence for such a transition associated to the onset of the centrifugal 
barrier was revealed in two HXRTs containing an X--ray
pulsar (V0332+53 with a spin period of 4.4 s, Stella, White \& Rosner
1986 and 4U 0115+63 spinning at 3.6 s, Tamura et al. 1992). 
Due to their relatively long spin periods, these X--ray pulsars are 
expected to have $\Delta\sim 400$. 
The transition to the propeller regime has probably been observed
during the outburst decay of the SXRT Aql X-1 (Zhang, Yu \&
Zhang 1998; Campana et al. 1998). 
Note that for short a neutron star spin period (2--4 ms as for the case of 
Aql X-1), the luminosity jump across the centrifugal barrier is expected to be 
much smaller (e.g. $\Delta=3$ for $P\sim 3$ ms) and the accretion 
luminosity in the magnetospheric accretion regime correspondingly higher. 
The transition to the propeller regime might be more gradual than 
assumed, due e.g. to a fraction of the inflowing material penetrating 
the magnetosphere at high latitudes and therefore accreting onto the neutron 
star surface (Stella, White \& Rosner 1986; Corbet 1996).

There are also uncertainties related to the extent  
of the luminosity jump associated  with the onset of the 
centrifugal barrier. Additional contributions to the luminosity on both 
sides of the centrifugal barrier might 
derive from the coupling of the magnetic field lines to the disk matter.
Priedhorsky (1986) pointed out that if the neutron star's magnetic field 
couples with matter in the inner disk beyond the magnetospheric
boundary, the rotational energy removed by spin-down torques from the neutron 
star can be released in the disk, increasing its luminosity.
Even in the absence of such a magnetic coupling, the propeller regime is 
probably characterised by the transfer of angular momentum and 
energy from the neutron star rotation to the inflowing matter, due to the 
azimuthal asymmetry of the magnetospheric boundary. 
The neutron star can therefore work as a flywheel, storing rotational energy
during the spin-up phases and releasing it in the disk during spin-down 
phases and likely also in the propeller regime.  
For fast rotating neutron stars 
these processes can lead to a substantially higher luminosity than that 
produced through accretion (see e.g. Priedhorsky 1986), therefore making 
the transition to the propeller regime more difficult to 
observe. 

Despite all these uncertainties, the onset of the propeller regime in a SXRT
might be revealed through a sudden spectral change or a steep decrease 
of the X--ray flux, as in the case of RossiXTE and 
BeppoSAX observations of Aql X-1 (Zhang, Yu \& Zhang 1998; Campana et al. 
1998).

In the propeller regime, matter accreting onto the magnetospheric 
boundary will in any case release its binding energy 
(see Eq. \ref{L_min}).
In the case of spherical accretion this luminosity amounts to 
\be
L^{sp}(r_{\rm m})\sim 10^{34}\,B_9^{-4/7}
\,M_{1.4}^{8/7}\,\mdot_{15}^{9/7}\,R_6^{-12/7} {\rm \ erg\,s^{-1}} 
\label{lrmsp}
\en
\noindent (In the case of disk accretion a luminosity by a factor of 
2 lower should result).
The luminosity in the disk accretion case is comparable to that of the 
spherical case, due to the similar extent of the magnetosphere for 
the same accretion rate (cf. Eq. \ref{compa}). 
In the propeller regime the fate of the infalling matter is 
uncertain: it can either be swept away by the magnetospheric drag 
at the expense of the rotational energy of the neutron star
(as a result of either a supersonic or a subsonic propeller; Davies \& 
Pringle 1981; Ghosh 1995) or accumulate outside the magnetospheric boundary
possibly giving rise to a quasi-steady atmosphere (Davies \& Pringle
1981). In the latter case, the accumulation of matter at the magnetospheric 
boundary can give rise to a pressure build-up, that might eventually push 
the magnetospheric boundary inside the corotation radius, therefore
leading to accretion onto the neutron star surface.

The steep luminosity decline that characterises the propeller phase in  
Aql X-1 is considerably faster than what would be expected extrapolating
the mass inflow rate from the first part of the outburst to the 
magnetospheric accretion regime (see Fig. 1). 
We note that the spectral transition accompanying the onset of the
centrifugal barrier may  modify the irradiation of the disk, therefore 
speeding up the end of the outburst (Campana et al. 1998).
Alternatively, if matter ejection takes place in the propeller phase,
its interaction with the inflowing matter can contribute in turning off
the outburst.

As the mass inflow rate decreases further, the magnetospheric radius
expands until it reaches the light cylinder radius ($r_{\rm lc}=c\,P/2\,\pi$, 
with $c$ the speed of light). Consequently, the neutron star
electromagnetic pressure thus changes from static ($\propto r^{-6}$) to
radiative\footnote{The fraction of the radiation pressure interacting
with matter has been estimated to be $\sim 1$ from studies of
bow-shock nebulae around radio pulsars (Kulkarni \& Hester 1988;
Cordes, Romani \& Lundgren 1993).} ($\propto r^{-2}$). 
The pressure of the inflowing material scales as $r^{-5/2}$ 
in the case of spherical symmetry; it can be shown that scales 
as $r^{\alpha}$ with $-7/2<\alpha<-51/20$ for a standard accretion disk
depending on the chosen region.
The radiative pressure of the radio pulsar is therefore characterised by a 
flatter radial dependence than the pressure of the inflowing matter. 
If the pressures balance at $r_{\rm lc}$, then for any larger radii the 
pulsar radiation pressure will dominate. 
Under these conditions, no stable equilibrium exists and the accreting
matter begins to be pushed outwards (Illarionov \& Sunyaev 1975; Shaham 
\& Tavani 1991). Matter inflowing through the Roche lobe is stopped by 
the radio pulsar radiation pressure, giving rise to a shock front. 

The limiting mass inflow rate, $\mdot_{\rm lc}$, below which the radio
pulsar mechanism will turn on, is determined by the condition 
$r_{\rm m}\sim r_{\rm lc}$. This corresponds to a minimum luminosity 
produced by accretion onto the magnetosphere of
\be
L_{\rm lc}=L(r_{\rm lc})=G\,M\,\mdot_{\rm lc}/r_{\rm lc} \label{L_lc}
\en
which in the spherical approximation becomes
\be
L_{\rm lc}^{sp}=7\times10^{31}\,B_9^2\,M_{1.4}^{1/2}\,
P_{-2}^{-9/2}\,R_6^{6} {\rm \ erg\,s^{-1}} \label{L_lc_sp}
\en
In the case of disk accretion a factor of $\sim 70$ lower luminosity 
is expected for the same set of parameters (see Eq. \ref{LlcGL}).
The observability of a radio pulsar signal during this phase is discussed 
in detail in the Section 5.3.1.
In this regime, a fraction of the neutron star spin-down 
luminosity can be emitted in the X--ray band as a result of the interaction 
between the relativistic pulsar wind and the incoming matter. 
A conversion efficiency of the spin-down luminosity up to $\sim 10\%$ can 
be expected in these systems (Tavani 1991; see also below). 

The mass inflow rate required to overcome the pulsar radiation barrier is 
higher than that for which the radio pulsar turns on, such that a limit cycle 
is expected. In fact, in order to overcome the pulsar radiation, the 
pressure of the matter inflow must dominate the radiation
pressure ($P_{\rm rad}=L_{\rm sd}/4\,\pi\,r\,c$, where $L_{\rm sd}$ is
the spin-down luminosity) at the shock radius, in the proximity of the 
inner Lagrangian point (where matter spills out with almost constant 
pressure, dominated by its thermal energy, e.g. Meyer \& Meyer-Hofmeister 
1983). 
Once the radiation barrier is won, the matter inflow can proceed inside
the light cylinder radius quenching the radio pulsar mechanism (e.g.
Illarionov \& Sunyaev 1975; Shaham \& Tavani 1991; Lipunov 1992).

The condition for the quenching the radio pulsar emission is not easy 
to derive as it depends on the geometry of the shock front. 
The most likely possibility is represented by the formation of an 
``outer bubble", supported internally 
by pulsar radiation pressure and continuously fed by mass loss from 
the companion (as in the case of PSR~1744--24A; Lyne et al. 1990; 
Tavani \& Brookshaw 1991, 1993). An accurate estimate of the mass needed 
to overcome the pulsar pressure requires detailed magneto-hydrodynamical 
simulations.

\begin{figure*}[htb]
\psfig{figure=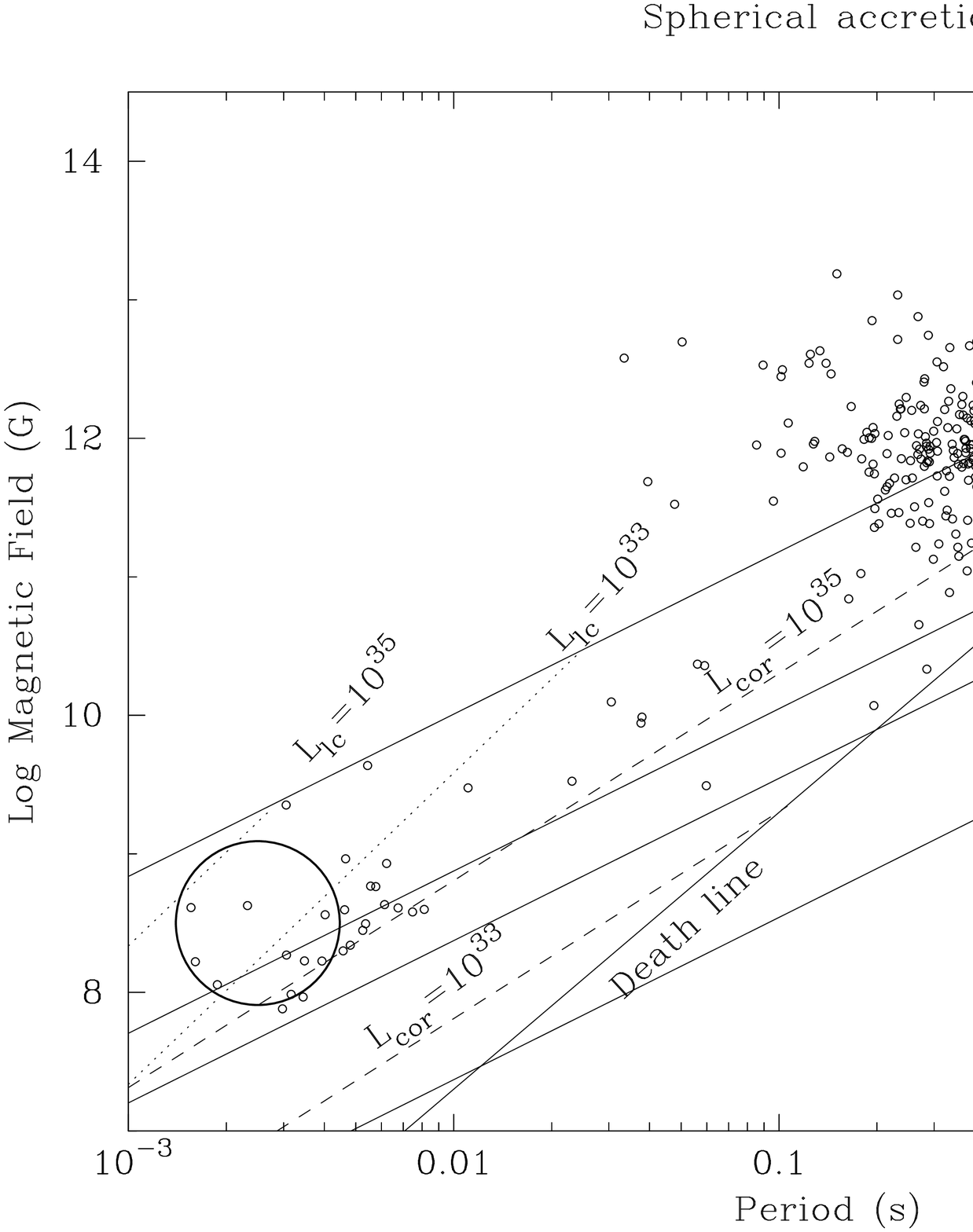,width=9cm}
\caption{Surface magnetic field plotted versus spin period of known
pulsars (circles), taken from Taylor, Manchester \& Lyne (1993). The
death line corresponds to the polar cap voltage below which the radio
pulsar activity switches off (e.g. Bhattacharya \& van den Heuvel 
1991). The death line is here described as $B_9 \simeq 2\times 10^{-2} 
\,P_{-2}^2\,R_6^{-3}~{\rm ~G}$ (Ruderman \& Sutherland 1975; 
Bhattacharya \& van den Heuvel 1991; Chen \& Ruderman 1993). 
However, Rudak \& Ritter (1994) and Phinney \& Kulkarni (1994) 
suggested to reshape the death line in the MSPs region to account 
for curvature radiation reaction. This results in a flatter death line 
for spin periods smaller than $\sim 0.1$ s, nearly coincident with 
the locus of pulsars with characteristic age $\tau_c=P/(2\,\dot{P})$ equal 
to the age of the Universe (10 Gyr, i.e. the so-called ``Hubble'' line; here
$\dot{P}$ is the spin period derivative). A different death line was suggested 
by Bj\"ornsson (1996). In this paper, however, we adopt the standard 
death line. For more details on the death line, the graveyard
and the ``Hubble'' line see Srinivasan (1989).
Solid lines represent the limit given by Eq. \ref{lminsp} (in erg~s$^{-1}$).
Note that the line for $L=L_{\rm Edd}=1.8\times
10^{38}\,M_{1.4}\ergs$ corresponds to the so called spin-up line.
Lines marking the maximum accretion luminosity emitted in the 
propeller regime (i.e. for $r_{\rm m}=r_{\rm cor}$) are indicated with dashes 
for selected luminosities (in $\ergs$ units).
Lines corresponding to the minimum accretion luminosity in the propeller 
regime (i.e. for $r_{\rm m}=r_{\rm lc}$) are depicted for selected
luminosities (measured in $\ergs$) with dots. 
Therefore, if a neutron star is in the propeller regime and produces a given
accretion-induced luminosity, then it must lie within the region
between the corresponding lines $L_{\rm cor}$ and $L_{\rm lc}$.
The large circle in the millisecond period-low magnetic field region 
includes the kHz QPO sources.}
\label{Fsph}
\end{figure*}
 
\section {X--ray emission in quiescence}
 
Several SXRTs have been detected during the quiescent phase at a level
of $L^q\sim 10^{32}-10^{33}\ergs$.
The X--ray spectra observed by ROSAT are usually soft and are 
characterised by a
black body temperature of $\sim 0.3$ keV or, equivalently, a power-law
slope of $\Gamma\sim 3$ (as detected in the 0.1--2.4 keV energy band).
Observations with larger energy bands (ASCA, BeppoSAX)
have provided evidence for the existence of a hard energy tail, in addition 
to the soft black body spectrum in Aql X-1 and Cen X-4 during their 
quiescence states.

The origin of this quiescent X--ray emission is still open to a number of
possibilities. Coronal activity of the companion star can be ruled out 
since late type low mass stars produce an X--ray
luminosity of $10^{32}\ergs$ at the most (Dempsey et al. 1995; Verbunt 1996). 
The quiescent emission in SXRT could result from:
 
\noindent 
(1) accretion onto the neutron star surface;\\
(2) accretion down to the magnetospheric
radius (when the ``centrifugal barrier" is closed, Paper I);\\
(3) non-thermal processes powered by the rotational energy loss of a
rapidly spinning neutron star (Tavani \& Arons 1997);\\
(4) thermal emission from the cooling neutron star.

\subsection {Accretion onto the neutron star surface}
 
In this regime, the requirement that 
the centrifugal barrier is open for $L_{\rm min}^{obs}=10^{33}\ergs$ 
translates into a fairly stringent condition on the neutron star
parameters, implying long spin periods and/or low magnetic fields (see
Eq. \ref{lminsp} and Fig. 3; Paper I; Verbunt 1996).  

For the neutron star spin period frequency of 2--4 ms inferred from recent
observations of Aql X-1, accretion onto the neutron star surface during
quiescence can take place only if $\lsim 5\times 10^6$ G. 
In this case then SXRTs could not be among the progenitors of MSPs. 
 
Models of the X--ray emission produced by a weakly magnetic neutron star
accreting (onto the polar caps) at $\sim 10^{33}\ergs$ predict a  
soft black body spectrum ($k\,T_{\rm bb}\sim 0.4$ keV) with a (weak) high 
energy tail due to Compton heating of thermal photons in the external 
part of the hot atmosphere (Zel'dovich \& Shakura 1969; Zampieri et al. 1995).  
Available X--ray spectra of the quiescent state of several SXRTs are 
consistent with a black body model with an emitting area a 
factor of $\sim 10^{-2}$ smaller than the neutron star surface, even if
the recent evidence for a hard tail in both Cen X-4 and Aql X-1 cannot 
easily accounted for by this model.

\subsection {Accretion down to the magnetospheric radius}
 
Another possibility is that the quiescent state of SXRTs is powered 
by accretion onto the magnetospheric boundary, such that an accretion 
luminosity of $L_{\rm m}=G\,M\,\mdot/r_{\rm m}$ is released (reduced by 
a factor of 2 in the likely case of disk accretion). 
This propeller regime is characterised by $r_{\rm {\rm cor}}\lsim 
r_{\rm m}\lsim r_{\rm {\rm lc}}$ (see Eq. \ref{L_lc}) and, therefore, 
the accretion luminosity released by the disk is $L_{\rm lc}\lsim 
L_{\rm rm} \lsim L_{\rm cor}$.
In Fig. 3, the lines of constant $L_{\rm lc}^{sp}$ and 
$L_{\rm cor}^{sp}$ are drawn. It should be emphasised that $L_{\rm rm}$ 
provides only a lower limit to the total quiescent luminosity, which might 
include additional contributions from the release of neutron star rotational
energy through the interaction between the rotating magnetosphere and the 
inflowing matter. 
Assuming that most of the quiescent luminosity $L^q$ of a SXRT results from
accretion onto the magnetosphere, the neutron star is constrained to lie 
in the region of the $B-P$ diagram between the lines $L^q=L_{\rm lc}$ and 
$L^q=L_{\rm cor}$. For $L^q=10^{33}\ergs$, this region is fairly wide 
and contains a large number of MSPs (see Fig. 3), even though it  
is relatively distant from the classical spin-up line. In this interpretation, 
the neutron stars of SXRTs cannot be the progenitors of the highest magnetic 
field MSPs. Note that the line $L_{\rm lc}= 10^{33}\ergs$ straddles the region
of the  $B-P$ diagram where neutron stars of LMXRBs are expected to lie, 
according to the beat-frequency interpretation of the kHz QPOs. 

Little theoretical work has been carried out so far on the spectrum emitted 
by a neutron star in the propeller regime. 
If the inflowing matter is in the form of an
optically thick accretion disk, most of the flux should be 
emitted in the UV/soft X--ray band (e.g. Shakura \& Sunyaev 1973).
If the magnetosphere remains small (as expected for low magnetic fields and 
spin periods) a hot, inner Comptonising corona might form in analogy with 
the case of persistent LMXRBs, giving rise to a power-law like spectrum.  
One might speculate that the formation of an inner disk corona
is due to the interaction between the disk and the
magnetosphere in the propeller regime (e.g. through shocked plasma).

In the case of Aql X-1 the onset of the propeller is marked by a 
steep increase of the hardness ratio (Zhang, Yu \& Zhang 1998). 
The BeppoSAX spectrum shows a soft component (modeled as a black body 
with $k\,T_{\rm bb}\sim 0.3$ keV) together with a power-law
with photon index $\sim 2$ extending up to $\sim 100$ keV 
(see Section 2.1.1 and Campana et al. 1998).

\subsection {Pulsar shock emission} \label{shock}
 
One can infer the magnitude of the X--ray emission that would be
expected in the case of pulsar-driven SXRTs nebulae in quiescence,
relying on the recent observations of the PSR~B1259--63 system and 
the theory of plerionic high-energy emission. As the observations of the
binary MSP PSR~B1259--63 demonstrate
(Tavani \& Arons 1997), non-thermal X--ray emission is produced by the
interaction of the relativistic pulsar wind with gaseous material.
ASCA (Kaspi et al. 1995) and Compton~GRO (Grove et al. 1995) observations of
the 47~ms pulsar PSR~B1259--63 near periastron (when the pulsar
interacts more strongly with the outflow from the Be star companion)
clearly show a power-law spectrum with $\Gamma \sim 1.6-2.0$ extending
from $\sim 1$~keV up to $\sim 200$~keV as well as absence of pulsations.
The high-energy emission near periastron cannot be due to accretion
onto the surface of the neutron star (Tavani \& Arons 1997) nor
to accretion down to the magnetospheric radius (Campana et al.
1996).  Rather it is in agreement with a synchrotron model of
plerionic-like high-energy shock emission expected for a pulsar wind
composed of electrons, positrons and possibly ions constrained to have
Lorentz factor $\gamma_1 \goe 10^6$ and plasma magnetisation $\Sigma$
(ratio of Poynting energy flux to the particle kinetic energy flux)
less than unity.  The PSR~B1259--63 observed broad-band efficiency of
conversion of pulsar spin-down energy into high-energy radiation is
(in the energy range 1--200 keV) $\varepsilon_{1259-63} \sim 1-3\%$
(Kaspi et al. 1995; Grove et al. 1995). Emission of
comparable efficiency is expected also to be radiated in that system
between 200~keV and 1--5~MeV (Tavani \& Arons 1997). 
The observed shock efficiency depends obviously on the geometric
characteristics of the pulsar cavity containing the pulsar wind. Since
the PSR~B1259--63 pulsar wind pressure probably disrupts the
equatorial Be star outflow by `breaking open' the disk-like outflow,
the overall efficiency may be lower than for a spherically symmetric
pulsar cavity (as in the case of the Crab Nebula, e.g. Kennel \&
Coroniti 1984). 

If the acceleration shock timescale is less than the cooling timescale,
the high-energy emission has a non-thermal nature (cooling is caused
by synchrotron emission and possibly inverse Compton scattering in the
optical background from the companion star).  In this case, the
power-law energy spectrum, calculated in an acceleration model based
on magnetosonic absorption of ion induced waves (e.g. Arons \& Tavani
1993), falls in the range $\epsilon_1 < \epsilon < \epsilon_m$, where
$\epsilon_1 = 0.3 \,\gamma_1^2\, \hbar\, \omega_{cyc}(B_2)$
(typically $\epsilon_1  \sim 1$~eV for $B_9/P_{-2}^3 \sim 1$; $\hbar$ is the
rationalised Planck's constant)
with $\omega_{cyc}$ the electron/positron cyclotron frequency and
\begin{eqnarray}
\epsilon_m & = &  \gamma_m^2 \,\hbar\, \omega_{cyc}(B_2) 
\simeq 35 \, \left(\frac{\eta}{0.3}\right)^2 \,v_{8}\, \nonumber  \\ & & 
\left[\frac{n_8}{(1+\Sigma)}\right]^{1/2}\, 
\si^{1/2}\,\frac{B_9}{P_{-2}^3} \   {\rm keV}
\label{eq:max-bubble-energy}
\end{eqnarray}
\noindent is the emission energy for the upper exponential cutoff of
the synchrotron spectrum.  In Eq. \ref{eq:max-bubble-energy} we use the notation 
$v_8 = v/(10^8 \cms)$ for the outflow velocity at the shock radius
distance, $n_8 = n/(10^8 \, \rm cm^{-3})$ for the nebular density and
$\eta$ for the efficiency of particle acceleration in the pulsar
magnetosphere as defined in Arons \& Tavani (1993). In 
Eq. \ref{eq:max-bubble-energy} it is assumed that a
(conservative) maximum-to-initial ratio of post-shock energies
$\gamma_m/\gamma_1 \sim 10$. The power-law index depends on the shock
acceleration injection mechanism and efficiency. Based on the analogy
with the PSR~B1259--63 system, we infer a photon spectral index $1.6
\lsim \Gamma \lsim 2$, for an index of the underlying post-shock energy
distribution function of the radiating pairs in the range $\sim 2-3$.

\begin{figure*}[htbp]
\psfig{figure=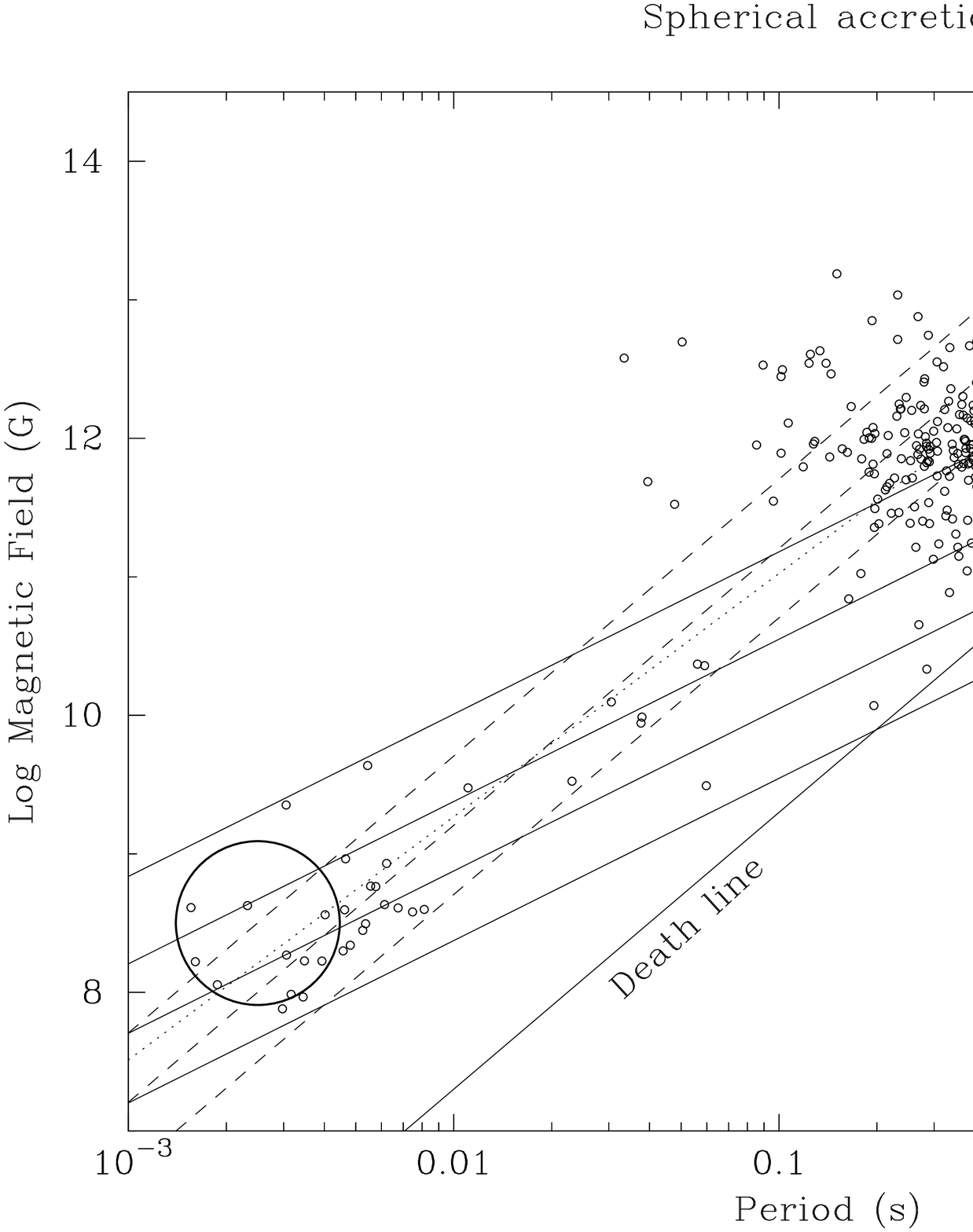,width=9cm}
\caption{Surface magnetic field plotted versus spin period of known
pulsars (circles). The Eddington spin-up line is shown as a continuous line as
well as other spin-up lines for selected luminosities.
Dashed lines mark selected values of the spin-down luminosity.
The dotted line separates the region in which the
neutron star in a SXRT is expected to undergo secular spin-down (left)
from that in which spin-up takes place (right) in the case of a radio
pulsar being active during the quiescent phase (see also text). 
The large circle in the millisecond period-low magnetic field region 
includes the kHz QPO sources.}
\label{Fls}
\end{figure*}

Substantial flux in the hard X--ray range should be observable for the
acceleration-dominated case especially in the case of fast spinning
neutron stars, and the spectral shape might provide an interesting
signature for the existence of enshrouded pulsars in SXRTs. In
particular power-law spectra are predicted, the photon index 
of which can be related to the characteristics of the shock front between the
radio pulsar and the incoming matter.

The luminosity budget of quiescent SXRTs can be accounted for by a 
radio pulsar only if it is spinning rapidly at a few millisecond level 
and has a magnetic field of at least $10^8$ G. Fig. 4 shows the 
lines of constant spin-down luminosity in the $B-P$ diagram.
Assuming a conversion efficiency of spin-down luminosity to X--ray of 
$\varepsilon \sim 1-10\%$, 
which is plausible for enshrouded pulsars (Tavani 1991), we can match 
the $\sim 10^{33}\ergs$ luminosity observed in quiescent SXRTs,
if their neutron stars lie in the region between $L_{\rm sd} = 10^{34}$ 
and $10^{35} \ergs$. 
Note that the low $B$ and $P$ section of this region overlaps quite well with 
the region inferred from the beat-frequency interpretation of the kHz QPOs 
in LMXRBs. 

For the pulsar shock emission regime to apply to the quiescent 
state of SXRTs, the transition to the propeller regime must take place 
at a relatively high luminosity. In particular, if the neutron star 
spin period is 2--4 ms and the magnetic field $\sim 10^8-10^9$ G 
(as indicated by kHz QPO sources), the transition to the propeller 
regime involves a small luminosity gap (a factor of a few; cf. Eq. \ref{corbet})
and is expected to take place at a luminosity level of 
$\sim 10^{35}-10^{37}\ergs$ (cf. Eq. \ref{lminsp}).

The outburst decay of Aql X-1 as observed with RossiXTE and BeppoSAX showed
a small jump in luminosity around $10^{36}\ergs$ followed by a sudden increase 
in the hardness ratio and steepening of the luminosity decay. These phenomena
likely indicate the onset of the centrifugal barrier (Zhang, Yu \& Zhang 1998;
Campana et al. 1998).
The quiescent emission of Aql X-1 has been investigated in detail with BeppoSAX.
The spectrum can be satisfactory fitted by a soft black body component 
plus a hard power-law with photon index of $1-1.5$ (Campana et al. 1998).
The quiescent spectrum is significantly different from the one observed 
after the closure of the centrifugal barrier (characterised by a power-law 
photon index of $\sim 2$), suggesting that another change in the powering 
mechanism occurred. The quiescent X--ray luminosity of $6\times 10^{32}\ergs$ is 
consistent with being powered by shock emission induced by the relativistic 
wind of the rapidly spinning neutron star. The required 0.1--10\% conversion 
efficiency of spin-down luminosity to X--rays is consistent with modeling 
and observations of enshrouded pulsars (Tavani 1991; Verbunt et al. 1996).

We note also the the ASCA spectrum of Cen X-4 shows a power-law 
component extending up to at least $\sim 5$ keV in addition to a soft 
component with an equivalent black body temperature of $\sim 0.3$ keV 
and comparable luminosity (Asai et al. 1996a), suggesting that also in this
case the shock emission mechanism could be at work.

\subsubsection{Observability of pulsed radio emission in SXRTs}

We address here the expected radio luminosity of the MSPs contained in SXRTs. 
Taking the empirical relation between the mean 
pulsed luminosity at 400 MHz radio pulsar magnetic field and spin 
period for MSPs (Kulkarni, Narayan \& Romani 1990), we derive a radio 
luminosity of $L_{\rm radio}\sim 40\, B_9^{2/3}\,P_{-2}^{-4/3}\, 
{\rm mJy\,kpc^{-2}}$. The minimum detectable radio flux increases fairly 
rapidly for decreasing periods. Moreover, the dispersion measure
will heavily limit the observations: the mean dispersion measure of MSPs is 
about 40 pc cm$^{-3}$, with only a few MSPs with a value higher than 100 
pc cm$^{-3}$.

The visibility of pulsed radio emission is made more difficult also 
by the presence of the matter spilling 
out of the first Lagrangian point into the Roche lobe of the compact object.
If an ``outer bubble'' supported by pulsar radiation pressure ensues
(as in the case of PSR~1744--24A; Lyne et al. 1990) radio pulsed signal 
could only be observed sporadically, if at all (Tavani 1991).
Hydrodynamic simulations aimed at modeling the shock front and the 
evolution of the enshrouding matter (Brookshaw \& Tavani 1995)
indicate that only for small inclination angles pulsed radio emission 
can propagate relatively undisturbed. 
In particular, the optical depth to free-free (bremsstrahlung) 
absorption caused by the material outflowing from the companion star 
can easily be much larger than unity (Brookshaw \& Tavani 1995).
On the contrary, if radiation pressure sweeps away the mass 
inflow all together (as in the case of PSR~1957+20; Fruchter et 
al. 1992), the region between the two stars remains almost ``clean", providing
favourable conditions for observing a MSP during the quiescent phase of a SXRT.
We therefore conclude that radio observations of SXRTs in quiescence may 
not reveal any flux (continuum or pulsed) and still be compatible with 
the existence of an underlying enshrouded pulsar.

\begin{figure}[!htbp]
\psfig{figure=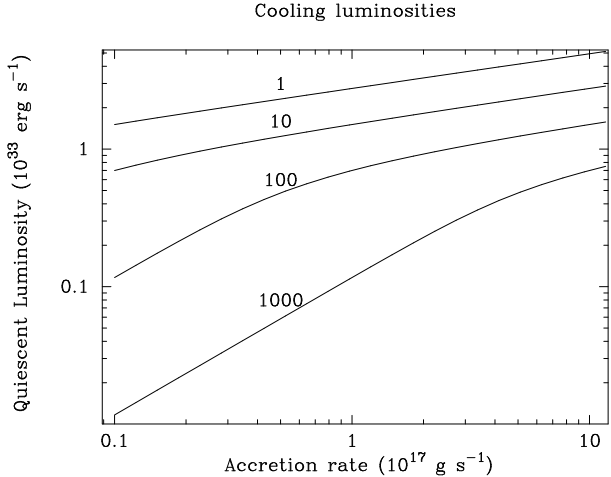,width=6cm}
\caption{Quiescent black body luminosities from a cooling neutron star 
in units of $10^{33}\ergs$ as a function 
of the time-average accretion rate during outburst in units of 
$10^{17}\gs$. From top to bottom, curves refer to values of the ratio 
of recurrence time to the duration of an outburst $\Theta=1,\,10,\,100,\,1000$, 
respectively.}
\label{monica}
\end{figure}

\subsection{Black body emission from the neutron star surface}

The quiescent X--ray flux could originate
from the release of thermal energy at the surface of a moderately hot
neutron star. The emission may result from radiative cooling of the
warm interior heated up during the accretion episodes giving rise to
the outbursts. This possibility is briefly investigated in Appendix B.

It is known that steady accretion at large enough accretion rates
(i.e. $\mdot\gsim 3\times 10^{14}\gs$) causes the neutron star interior to
increase its temperature
owing to the release of energy by stable hydrogen burning ignited at
the base of the envelope (Fujimoto et al. 1984; Miralda-Escud\'e,
Paczy\'nski \& Haensel 1990). On a relatively short time scale of
$t_{c,\,eq}\sim 2\times 10^4
\,\mdot_{17}^{-3/4}\,\chi_{-3}^{-3/4}\,M_{1.4}^{3/4}\ \rm {yr}$
(where $\chi=\chi_{-3}\,10^{-3}$ is the nuclear burning efficiency),
the equilibrium between nuclear heating and neutrino cooling results in a
core temperature $T_{c,\,8}^{eq}\sim 2.4\times
10^8\,\mdot_{17}^{1/8}\,\chi_{-3}^{1/8}\,M_{1.4}^{-1/8}$.
Radiation from the warm interior released at the surface 
could then be revealed when accretion stops.
Since a temperature gradient is established across the envelope, thermal 
radiation at a temperature of $T_{s}\sim 1.4\times 10^6\,
\mdot_{17}^{1/12}\,\chi_{-3}^{1/12}\,M_{1.4}^{-1/12} ~\rm {K}$ is
emitted, giving rise to a black body luminosity of
$L_{s}\sim 3\times
10^{33}\,R_6^2\,\mdot_{17}^{1/3}\,\chi_{-3}^{1/3}\,M_{1.4}^{-1/3}\ergs$
(Gudmundsson, Pethick \& Epstein 1983).  

This estimate provides only an {\it upper limit} since the
(possible) presence of a pion condensate in the star's interior reduces
considerably the core temperature and, in turn, the temperature of the
thermal radiation emitted at the stellar surface.  
In addition, owing to the intermittent character of accretion in a SXRT, 
the time necessary to attain equilibrium is
$\sim t_{c,\,eq}\times \Theta$ (where $\Theta$ is the
ratio of the recurrence time to the duration time of a SXRT outburst); this 
time can exceed the time over which heat leaks out through the envelope.
This would further  reduce  the
upper limit on $T_{c,\,8}^{eq}$ and in turn the black body luminosity
$L_{s}$ observed in quiescence.
In Fig. 5, $L_{s}$ is plotted against the mean
accretion rate during outburst, for different values of $\Theta$.
For $\Theta \lsim 100$ the luminosity is in the range $3\times
10^{32} - 2 \times 10^{33}\ergs$, which is comparable to the
range of luminosities observed during quiescence. 

Neutron star cooling is attractive in consideration of the soft X--ray spectra 
observed in quiescent SXRTs. However, the most reliable determinations 
of the equivalent black body radii hint to smaller values than 
the standard $\sim 10^6$ cm neutron star radius. Moreover, the ROSAT HRI 
observation of a flux variation on a timescale of a few days in the quiescent 
flux of Cen X-4 (Campana et al. 1997) as well as the hard tails observed in 
Aql X-1 and Cen X-4, can hardly be reconciled with quiescent emission
being powered by this mechanism alone.
Note that thermal emission from neutron star cooling does not impose 
any relevant restriction on $B$ and $P$. Therefore the possibility that
this mechanism contributes with a soft and nearly steady 
black body like component to the quiescent emission of SXRTs remains open.

\section {Spin period evolution}

Different spin period histories for the neutron stars in SXRTs are possible
depending on which mechanism is operating during quiescence.
In the following discussion we neglect the secular decrease of the 
magnetic field that possibly results from accretion.
If a radio pulsar is active during
quiescence, then the spin period evolution is determined by the
competition between the disk spin-up accretion during
the outbursts (with a torque of $N=\mdot\,\sqrt{G\,M\,r_{\rm m}}$) and  
spin-down caused by radio pulsar losses in
the quiescent phases; these two effects balance for 
\begin{eqnarray}
B_9 & \simeq & 2\times 10^{-2} \,
P_{-2}^{187/80}\,L_{37}^{41/80}\,M_{1.4}^{-91/320} \nonumber \\
 & &\,R_6^{-199/80}\,\Theta_{100}^{-187/320}
\label{peqpuls}
\end{eqnarray}
\noindent where $L_{37}$ is the time-averaged outburst luminosity in
units of $10^{37}\ergs$ and $\Theta_{100}$ the ratio between the
time spent in quiescence and the average outburst duration in units of 100.
SXRTs above the corresponding line in the $B-P$ diagram (see Fig.
4) would undergo secular spin-down, whereas a secular 
spin-up would characterise SXRTs below the line. 
The timescale for the period evolution is $\tau \equiv P/\pdot\sim 
10^{10}\,B_9^{-2} \,P_{-2}^2$ yr in the case of a radio pulsar spin-down, 
and $\tau\sim 10^9 \,B_9^{-2/7}\,M_{1.4}^{3/7}\,L_{37}^{-6/7}\,P_{-2}^{-1}$ 
yr (e.g. Henrichs 1983) in the case of disk-like accretion for the same 
set of parameters as above.
Note that the position of the equilibrium line in Fig. 4 agrees
well both with the enshrouded pulsar regime described in Section 5.3 
and with the region suggested by the kHz QPOs.

If the accreting matter is stopped at the magnetospheric radius during
the quiescent phase, the neutron star enters the propeller regime and a
pronounced spin-down results that will affect its period evolution.
Different spin-down mechanisms have been proposed to describe the action
of the centrifugal barrier.
The spin-down torque predicted by the Illarionov \& Sunyaev (1975) 
propeller mechanism is the same in module as for the spin-up by a disk 
(i.e. the torque $N=-\mdot\,\sqrt{G\,M\,r_{\rm m}}$). However, the system 
spends a factor $\Theta$ more time in the propeller 
regime with an accretion rate much smaller.
In this case the equilibrium does not depend on the neutron star parameters
(as long as $B$ and $P$ are such that the neutron star remains in the propeller
regime during quiescence).
A net spin-down will occur if 
\be
L_{37}\lsim 0.6\,(L_{33}^q)^{7/9}\,B_9^{4/9}\,M_{1.4}^{1/9}\,R_6^{1/3}
\,\Theta_{100}^{7/6}
\label{pp}
\en
\noindent (where $L_{33}^q$ 
is the quiescent luminosity in units of $10^{33}$ erg s$^{-1}$).
A spin-up will result if the condition above if not satisfied. 
The spin-up/down time scale in this regime is $\tau\sim 10^7 \,
B_9^{-2/7}\,M_{1.4}^{3/7}\,L_{37}^{-6/7}\,P_{-2}^{-1}$ yr (e.g. Henrichs 1983),
which is much shorter than in the case of a radio pulsar.
Different versions of the propeller regime have been proposed.
The (only) one relying on numerical simulations provides a much more 
efficient spin-down  mechanism (Wang \& Robertson 1985). 
Milder versions of the propeller mechanism have also been suggested 
(e.g. the supersonic propeller of Davies \& Pringle 1981; Lipunov \& 
Popov 1995). 
Though still highly uncertain, the modeling of the propeller
effect suggests that substantial period variations of either sign
can take place on relatively short timescales if quiescent SXRTs 
are in this regime.

\section{Conclusions}

The properties of SXRTs in outburst are clearly linked to those of 
persistent LMXRBs, as testified by e.g. their X--ray spectra, 
the occurrence of type I bursts, optical brightenings and kHz QPOs,
indicating that SXRTs host weakly magnetic neutron stars.

For X--ray luminosities much below the outburst maxima, SXRTs show similarities 
with transient BHCs. In particular, as the luminosity decreases at a level 
of $\sim 10^{36}-10^{37}\ergs$ in a few SXRTs (as well as in persistent LMXRBs)
the X--ray spectrum shows a transition from a relatively 
soft thermal spectrum to a power-law like spectrum extending up to 100 keV
(Barret \& Vedrenne 1994; Mitsuda et al. 1989; Harmon et al. 1996). 
A similar spectral transition occurs also across the ``high'' and ``low 
states'' of several BHCs (e.g. Tanaka \& Shibazaki 1996). 
The spectral hardening in BHCs has been interpreted in terms of
a change in the disk to an advection-dominated regime (Narayan \& Yi 1995).
This regime has also been invoked to explain the low luminosity emission 
of the BHCs A~0620--00 (Narayan, McClintock \& Yi 1997) and V404 Cyg 
(Narayan, Barret \& McClintock 1997). 
We note however that in the case of SXRTs the presence of a ``hard surface''
(either the star surface or the magnetosphere) makes the application of simple 
advection-dominated models questionable.

Also in quiescence, SXRTs and transient BHCs show similar properties, 
displaying a soft spectral component (with equivalent black body temperatures 
of $\sim 0.2-0.3$ keV; e.g. Tanaka \& Shibazaki 1996).
However ASCA and BeppoSAX observations of Cen X-4 (Asai et al. 1996a) and Aql X-1
(Campana et al. 1998) during quiescence demonstrated the presence
of hard tails which have not been detected in BHCs.

We note also that the Aql X-1 outburst decay resembles closely the 
evolution of dwarf novae outbursts, with a drastic turn off of the luminosity 
(e.g. Osaki 1996). Models of low mass X--ray transient outbursts hosting an old
neutron star or a black hole are largely built in analogy with dwarf
novae outbursts. In particular, van Paradijs (1996) showed that the different 
range of time-averaged mass accretion rates over which the dwarf nova and 
low mass X--ray transient outbursts were observed to take place is well 
explained by the higher level of disk irradiation caused by the higher accretion 
efficiency of neutron stars and black holes.
However, the outburst evolution of low mass X--ray transients
presents important differences. In particular, the steepening in the X--ray
flux decrease of Aql X-1 has no clear parallel in low mass X--ray transients
containing BHCs. The best sampled light curves of these sources show an 
exponential-like decay (sometimes with a superposed secondary outburst) 
with an $e-$folding time of $\sim 30$ d and extending up to four decades in 
flux, with no indication of a sudden steepening (Chen, Shrader \& Livio 1997). 
In addition, 
BHC transients display a larger luminosity range between outburst peak and 
quiescence than neutron star SXRTs (Garcia et al. 1998 and references therein).
Being the mass donor stars and the binary parameters quite similar in the
two cases, it appears natural to attribute these differences to the
different nature of the underlying object.
In principle after the decay of a neutron star SXRT outburst the X--ray emission 
might be dominated by different mechanisms, notably accretion
onto the magnetosphere, emission from an enshrouded MSP or cooling from
the neutron star surface. None of these has an equivalent in the 
case of BHCs.

In the case of Aql X-1 for which a spin period frequency of 2--4 ms has been 
inferred, accretion onto the neutron star surface during the quiescent state of 
SXRTs would imply a neutron star magnetic field $\lsim 5\times 
10^6$ G.  Extrapolating this result, one would conclude that SXRTs could not 
be among the progenitors of MSPs. 
Cooling of the neutron star surface can easily account for  
the softness of X--ray spectra. However, the observation of a flux variation 
on a timescale of a few days in the quiescent flux of Cen X-4 
(Campana et al. 1997) can hardly be reconciled with this mechanism alone.
Moreover, the observation of a harded spectral component in Aql X-1 (Campana 
et al. 1998) and Cen X-4 (Asai et al. 1996a) poses problems to accretion onto 
the surface and cooling mechanisms.

Accretion onto the neutron star magnetosphere is one interesting possibility
to explain the quiescent emission of SXRTs. If this regime applies, the allowed
region in the $B-P$ diagram, even if far from MSPs near the 
Eddington spin-up line, contains a relevant number of MSPs and straddles 
the region where neutron stars of LMXRBs are expected to lie.
The spin period evolution in this case is mainly dictated by the 
ratio of the mean luminosity in outburst and during quiescence, so that either
spin-up and spin-down are possible, even if a secular spin-down is 
more likely (cf. Eq. \ref{pp}).

Shock emission powered by an underlying MSP is the other plausible mechanism
for the quiescent emission of SXRTs.
In this regime, SXRTs would occupy a region in the $B-P$ diagram containing  
a large number of MSPs and likely including kHz QPO sources. 
The equilibrium line between spin-up during outburst and spin-down due to 
radiative losses in quiescence, lies just in the middle of the expected 
neutron star spin-down luminosities.
BeppoSAX observations of Aql X-1 provide the best evidence so far for this 
mechanism (Campana et al. 1998). The predicted spectrum show a substantial 
flux at energies beyond the soft (e.g. ROSAT) energy range. 
However, a soft component is present in all SXRTs detected in quiescence, so 
that other components have to be invoked, like a more complex shock emission 
mechanism or the contribution from the cooling neutron star.

For the pulsar shock emission regime to apply, the 
closure of the centrifugal barrier must take place at a level of 
$\sim 10^{36}\ergs$ (cf. Eq. \ref{lminsp}), with a small jump 
in luminosity (cf. Eq. \ref{corbet}), as observed in Aql X-1. 
This is also the range of luminosities characterising the spectral 
hardening observed in persistent LMXRBs and SXRTs. This spectral 
change has been indeed related to the transition to the propeller regime
in the case of Aql X-1 (Zhang, Yu \& Zhang 1998; Campana et al. 1998).
In this interpretation SXRTs represent the immediate progenitors of MSPs. 

Estimating the current number of SXRTs in our Galaxy is not an easy task, 
due to the presence of strong selection effects (e.g. large column densities 
along the galactic plane; lower peak luminosities with respect to BHCs; 
limited temporal coverage, etc.). A likely number is a few hundreds for mean
recurrence time of $\sim 10$ yr and about a thousand for a recurrence 
time of $\sim 50$ yr (Tanaka \& Shibazaki 1996).

Despite the recent discoveries on Aql X-1 by Ros\-siXTE and BeppoSAX, 
deeper studies are necessary to better assess the nature of the
neutron stars in SXRTs. The case of Aql X-1 emphasises the importance of 
combining the nearly continuous monitoring of the outburst evolution that 
can be obtained by large field X--ray instruments, with deeper and more 
detailed pointed observations by narrow field X--ray telescopes during 
the crucial phases of the outbursts.

\medskip
{\bf Note added in proof.}

After this paper was accepted for publication, we became aware that
in April 1998 RXTE revealed a transient X--ray source at a position
consistent with SAX J1808.4--3658 (in't Zand et al. 1998), a variable
X--ray burster in the direction of the galactic bulge.
During pointed RXTE observations highly significant coherent pulsations
at $\sim 401$ Hz were detected. These allowed also to measure an orbital
period of $\sim 2$~hr and a mass function of $\sim 4\times 10^{-5}~\msole$
(Wijnands \& van der Klis 1998; Chakrabarty \& Morgan 1998).

These results show that SAX J1808.4--3658 is a LMXRB, the first to show
coherent millisecond pulsations in its persistent emission. A radio pulsar
(perhaps a partially eclipsing one) might turn on after the X--ray outburst
ends. In any case SAX J1808.4--3658 further strengthens the link between
SXRTs and MSPs.

\medskip
{\small
{\it Acknowledgement.}
We thank an anonymous referee for providing useful comments. This work was 
partially supported through ASI grants.}

\appendix

\begin{figure*}[htb]
\psfig{figure=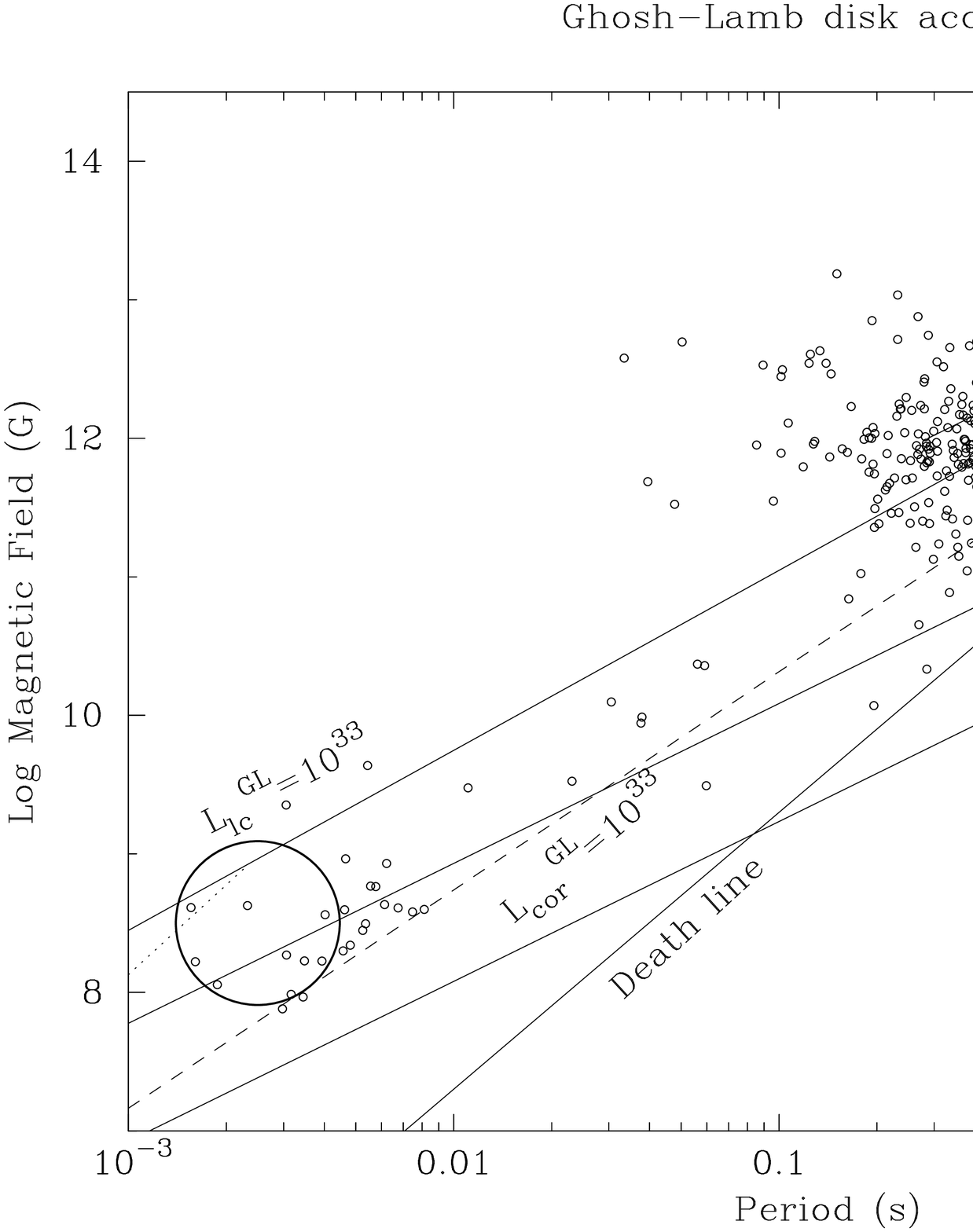,width=9cm} 
\caption{Surface magnetic field plotted versus spin period of
known pulsars (circles). 
Solid lines represent the spin-up lines for threaded-disk accretion 
given by Eq. \ref{LXGL}. The line corresponding to the Eddington luminosity 
has been calculated either for the gas pressure-domi\-na\-ted (lower part)
and for the radiation pressure-dominated (upper part) threaded-disk. 
The dotted lines indicate the maximum luminosity that can be released 
in the propeller regime (cf. $L^{GL}_{\rm cor}$) and the dashed line
the minimum luminosity (cf. $L^{GL}_{\rm lc}$). 
The large circle in the millisecond period-low magnetic field region 
includes the kHz QPO sources.}
\label{FGL}
\end{figure*}

\section{Ghosh and Lamb threaded-disk model}

In Section 4 we considered the case of spherical accretion, however an 
accretion disk most likely forms during SXRT
outbursts as a consequence of the high specific angular momentum of the
accreting matter.
Several models have been proposed for the interaction
between the accreting disk matter and the magnetosphere of a neutron
star (White \& Stella 1988). Here we consider the model described by 
Ghosh \& Lamb (1992 and references therein, hereafter GL92). 
In this model the neutron star magnetic field penetrates the disk over a
relatively large range of radii and two regions can be distinguished: an
outer transition zone, with a small coupling between matter and magnetic
field, and a narrow inner zone with the character of a boundary layer, in
which matter is brought to corotation by the magnetic field. GL92 adopted a
time-averaged description, taking a standard Shakura-Sunyaev disk
(Shakura \& Sunyaev 1973) and introducing an effective electrical
conductivity in order to describe the magnetohydrodynamic interaction
between matter and magnetic field. 

Solving a set of radial and vertical disk structure equations,
together with Maxwell's equations, GL92 derived the magnetospheric radius 
in a gas pressure-domi\-na\-ted accretion disk:
\begin{eqnarray}
r_{\rm m}^{GL} & = & 5\times 10^6\,\mdot_{15}^{-46/187}
B_9^{108/187}\, \nonumber \\
& &M_{1.4}^{-41/187}\,R_6^{324/187} {\rm ~cm} \label{rmGL}
\end{eqnarray}

In order to account for the torque exerted by magnetic field lines
onto the accreting matter, Ghosh \& Lamb (1979a; 1979b) introduced
the ``fastness'' parameter $\omega=\Omega_*/\Omega_k(r_{\rm m})$, with
$\Omega_*$ the angular frequency of the neutron star and $\Omega_k(r_{\rm
m})$ the angular frequency of the inner edge of the disk. We consider
here as the condition for the location of the centrifugal barrier $r_{\rm m} 
\lsim r_{\rm cor} \,(\omega_c)^{2/3}$, where $\omega_c$ is the critical 
value of $\omega$ for which the torque exerted by the 
accretion disk on the neutron star vanishes.

Different values have been proposed for $\omega_c$ (GL92; Wang 1995, 1996). 
Following Li \& Wickramasinghe (1996) we adopt $\omega_c\sim 0.5$.  With
this prescription, the minimum accretion-induced luminosity onto the
neutron star surface can be written as: 
\begin{eqnarray}
L_{\rm min}^{GL} & \sim & 10^{35}\,B_9^{54/23}
\,M_{1.4}^{-86/69}\, \nonumber \\
& & R_6^{139/23} \,P_{-2}^{-187/69}\,
\omega_{c,0.5}^{-187/69} \ergs\label{LXGL}
\end{eqnarray}
\noindent where also the factor of 1/2 has been included (see Section 4).
When the centrifugal barrier closes a decrease in the 
luminosity is predicted down to 
\begin{eqnarray}
L^{GL}(r_{\rm cor}) & \sim & 9\times 10^{33}\,B_9^{54/23}\,M_{1.4}^{-109/69}\, 
\nonumber \\ & & R_6^{162/23}\,P_{-2}^{-233/69}\,\omega_{c,0.5}^{-187/69} 
\ergs 
\label{LcorGL}
\end{eqnarray}
\noindent (compare this value with the one derived in the spherical
free-fall approximation, cf. Eq. \ref{lminn}).
If the accretion rate decreases further the magnetospheric radius 
can become comparable to the light cylinder radius. If this happen the
radio pulsar mechanism can turn on and start sweeping the infalling
matter away. In the case of an accretion disk with a boundary layer 
modeled following GL92, this occurs at a luminosity of
\begin{eqnarray}
L^{GL}(r_{\rm lc}) & \sim & 9\times 10^{29}\,B_9^{54/23}\,M_{1.4}^{5/46}\,
R_6^{162/23} \nonumber \\ & & P_{-2}^{-233/46} \,\omega_{c,0.5}^{-187/69} \ergs 
\label{LlcGL}
\end{eqnarray}

We have described only the gas pressure-dominated case in the region $b$ 
of a Shakura-Sunyaev disk, where the Thomson opacity dominates
over the free-free one. Ghosh \& Lamb have not worked out the case of the 
region $c$, whereas the case of the radiation pressure-dominated region $a$, 
the magnetospheric radius provided in GL92 is
\be
r_{\rm m}^A=3\times10^6\,B_9^{20/39}\,\mdot_{17}^{-2/13}\,M_{1.4}^{-5/39}
\,R_6^{20/13} {\rm ~cm}
\en

The relevant luminosities can then be easily derived:
\be 
L_{\rm min}^{GL,A}=6\times10^{35}\,B_9^{10/3}
\,M_{1.4}^{-2}\,R_6^{9}\,P_{-2}^{-13/3}\,
\omega_{c,0.5}^{-13/3} \ergs\label{LXGLA}
\en

\begin{eqnarray}
L^{GL,A}(r_{\rm cor})&=&4\times10^{34}\,B_9^{10/3}\,M_{1.4}^{-7/3}\,\nonumber \\
& &R_6^{10}\,P_{-2}^{-5}\,\omega_{c,0.5}^{-13/3} \ergs 
\label{LcorGLA}
\end{eqnarray}
\begin{eqnarray}
L^{GL,A}(r_{\rm lc})&=&5\times10^{28}\,B_9^{10/3}\,M_{1.4}^{1/6}\,\nonumber \\
& &R_6^{10}\,P_{-2}^{-15/2} \,\omega_{c,0.5}^{-13/3} \ergs 
\label{LlcGLA}
\end{eqnarray}

\subsection {A comparison between spherical and disk accretion models}

As apparent from Eqs. \ref{rmsp} and \ref{rmGL} the accretion flow is
dominated by the neutron star magnetic field within a radius, $r_{\rm m}$, 
the magnitude
of which depends sensitively on the geometry of the inflow. The ratio
of the magnetospheric radius in the spherical and disk geometry as a
function of the accretion rate is given by
\be
{{r_{\rm m}^{sp}} \over {r_{\rm m}^{GL}}}\sim 3\,\mdot_{17}^{-52/1309}\,
B_9^{-8/1309}\,M_{1.4}^{100/1309}\,R_6^{-24/1309}
\label{compa}
\en
\noindent As a consequence of this the minimum accretion-induced
X--ray luminosity in the spherical case exceeds the corresponding
value in the disk case by a factor of $\sim 10$ (for $B=10^9$ G and $P=10$ ms). 
In turn this causes the lines of constant minimum luminosity $L_{\rm min}$ in the
disk case to move upwards in the $B-P$ diagram, allowing for higher
magnetic fields and smaller spin periods. This is clearly seen in
Figs. \ref{Fsph} and \ref{FGL}. If a radio pulsar were to be present
in a SXRT at the end of disk accretion such radio pulsar can have a
higher spin-down luminosity.

Application of the standard beat-frequency model (Alpar \& Shaham 1985) to
bright X--ray pulsars led Wang (1995) to conclude that the neutron star's
dipole field fully or almost fully threads the disk. This would imply 
that the magnetospheric radius obtained in the case of spherical accretion  
is very similar to the one for disk accretion.

\section{Cooling neutron star model}

In this Appendix, we describe a simple model for the thermal evolution
of a neutron star in SXRTs.
Our aim is to determine the core temperature $T_c$ (and in turn the
surface radiation temperature $T_s$) of the neutron star heated up by
nuclear energy deposition that takes place during the accretion episodes 
that give rise to the outbursts.
First, we verify that  the evolution equation for $T_c$ reproduces
known results and later estimate how recurrent event of accretion
modify $T_c$ and the corresponding black body luminosity.
In the neutron star, we distinguish a uniform core at nuclear
densities and an envelope that surrounds it:  energy is exchanged
across the two regions and such a flow determines the equilibrium core
temperature and $T_s$, as described below.

The core is characterised by a single temperature $T_c$, owing to the
high conductivity of degenerate neutrons. $T_c$ evolves according
to the equation 
\be
C_c \,{dT_c\over dt}=L_N-L_{\nu}-Q \label{B1}
\en
\noindent where $C_c\sim 2\times 10^{38}\,M_{1.4}\,T_{c,8} \ \rm {
erg\,K^{-1}}$ is the heat capacity of the core and $T_{c,8}$ is its 
temperature in units of $10^8$ K.
In Eq. \ref{B1}, $L_N$ is the energy released per unit time by stable
hydrogen burning ignited at the base of the envelope.  At this
boundary a steep inversion of the temperature gradient arises, causing
the energy produced to flow into the core.  As a consequence of this effect
(see Fujimoto et al. 1984 and Miralda-Escud\'e, Paczy\'nski \& Haensel
1990 for details) $L_N$ enters Eq. \ref{B1} as the source 
of heat for the core, and can be estimated as
\be
L_N\sim \chi\,\mdot\,c^2 \sim 6\times 10^{35}\,\chi_{-3}\,\mdot_{17}
\ergs \label{B2} 
\en
\noindent where $\chi=\chi_{-3}\,10^{-3}$ is the efficiency of nuclear 
burning and $\mdot_{17}$ is the accretion rate in units of
$10^{17}\gs$.
Neutrino emission enters instead as a cooling term and can be
estimated by considering  nucleon-nucleon scattering as the dominant process 
(modified URCA)  
\be
L_{\nu}\sim 7\times 10^{31}\,T^8_{c,8}\,M_{1.4}\ergs
\label{B3}
\en
Additional neutrino losses occur if a pion condensate is present in
the core. $Q$ is a cooling term representing the energy flux
transferred outwards from the core to the surface via electron
conduction across the stellar envelope. Following Gudmundsson, Pethick 
\& Epstein (1983) we approximate $Q\sim 4\times 10^{16}\,T_{c,8}^2 \ergs$.

During thermal evolution, each process influences the changes in the
temperature on a characteristic time scale $\tau\sim T/{\dot T}$.  In
the core we are led to distinguish three time scales associated to
nuclear heating, neutrino and radiation cooling, respectively. We have
\be
\tau_N\sim {C_c\,T_c \over L_N}\sim 7\times 10^3\,
T^2_{c,8}\,\mdot_{17}^{-1}\,
\chi_{-3}^{-1}\,M_{1.4}\ \rm {yr} \label{B4}
\en
\be
\tau_{\nu}\sim {C_c \,T_c\over L_{\nu} }\sim 8\times 10^{6}
\,T_{c,8}^{-6}\ \rm {yr} \label{B5}
\en
\be
\tau_{k,core}\sim {C_c\,T_c \over Q}\sim 10^6\ \rm {yr} \label{B6}
\en

In the absence of accretion ($\mdot=0$), Eq. \ref{B1} reproduces the 
evolution of an isolated neutron stars cooling by neutrino and photon 
emission. The
energy flux across the envelope fixes the surface temperature $T_s$
which is determined uniquely by $T_c$; the resulting value is
$T_{s,6}\sim 0.9\,T_{c,8}^{1/2}$ and corresponding to a black body 
luminosity of
\be
L_s\sim 7\times 10^{32}\,T^4_{s,6}\ergs 
\label{B7}
\en
\noindent where $T_{s,6}$ denotes the surface temperature in units of
$10^6$ K.

In the case of steady accretion, nuclear reaction are ignited 
stably and the core  heats on the time scale $\sim \tau_N$. 
Thermal equilibrium is attained approximately when 
$L_{\nu}\sim L_N$ yielding an equilibrium temperature 
\be
T_{c,\,8}^{eq}\sim 2\,\mdot_{17}^{1/8}\,\chi_{-3}^{1/8}\,M_{1.4}^{-1/8}
\label{B8}
\en
This value is close to the one found by Fujimoto et al. (1984) and
Miralda-Escud\'e, Paczy\'nski \& Haensel (1990) using a more realistic model.
If there 
exists a pion condensate in the interior, the equilibrium temperature would
attain a lower value due to the enhanced neutrino energy losses that
occur at a rate $L_{\nu,\pi}\sim 2\times
10^{40}\,T_{c,8}^6\,M_{1.4}\ergs$. The inferred equilibrium
temperature is in this case $T_{c,\,8}^{eq}\sim 0.13\,\mdot_{17}^{1/6}\,
\chi_{-3}^{1/6}\,M_{1.4}^{-1/6}$.

If at the onset of accretion $T_c\ll T_{c}^{eq}$, the time necessary
for the core to attain equilibrium can be determined by integrating
the equation ${\dot T}_c=T_c/\tau_N$, having neglected both cooling
terms. The solution for $T_{c}(t)\propto t^{1/2}$ gives a
characteristic time for equilibrium 
\be
t_{c,\,eq}\sim 2\times 10^4\,\mdot_{17}^{-3/4}\,\chi_{-3}^{-3/4}\,
M_{1.4}^{3/4} \ \rm {yr} \label{B9}
\en

However, our neglect of $Q$ is justified only if the time for
radiative cooling through the envelope $\tau_{k,core}$ is much longer
than $t_{c,\,eq}$, a condition fulfilled when the 
steady  accretion 
rate (averaged over the quiescence time)
$\mdot_{17}>3\times 10^{-3}\,\chi_{-3}^{-1}\,M_{1.4}.$
In this case the envelope behaves mainly as an ``insulating" layer.  This
estimate is close to the limiting accretion rate below
which no stable hydrogen burning can occur (Miralda-Escud\'e, Paczy\'nski 
\& Haensel
1990) and, hence, the deposition  of energy into the core is
completely negligible.

In a SXRTs where accretion is intermittent hydrogen burning can occur
only during outbursts. In this case, energy is injected into the core
during the sporadic events of accretion. Equilibrium is therefore
attained on a time scale that can exceed  $\tau_{k,\,core}$. The leakage
of heat across the envelope may result in a much lower equilibrium
temperature, and in turn a much lower black body
luminosity.

Since $\tau_{k,core}$ is much longer than the expected recurrence time
between outbursts we can estimate approximately $T_{c}^{eq}$ solving
Eq. \ref{B1} for a time averaged accretion rate $\mdot \sim \mdot_{SXRT}/
\Theta$,
where $\mdot_{SXRT}$ is the mean accretion rate during outburst.
For $\Theta \gsim\,$100,
the equilibrium time is found to increase significantly
from the value for steady accretion 
of $10^4$ yr to $10^{7}$ yr, 
because of the leakage during quiescence.

\end{document}